\documentclass[aps, prl, twocolumn, amsmath,superscriptaddress,showkeys,showpacs]{revtex4-2}

\usepackage{graphicx}
\usepackage{graphicx,epstopdf}
\usepackage{dcolumn}
\usepackage{amsmath} 
\usepackage[colorlinks=true,linkcolor=blue,citecolor=blue]{hyperref}%
\usepackage{fontenc}
\usepackage{float}
\usepackage{amsthm}
\usepackage{subfigure}
\usepackage{color}
\usepackage{ragged2e}
\usepackage{bm}
\usepackage{enumerate}
\usepackage{hyperref}
\usepackage{xr}
\def\be{\begin{equation}}
\def\ee{\end{equation}}
\def\bea{\begin{eqnarray}}
\def\eea{\end{eqnarray}}
\def\bfg{\begin{figure}[H]}
\def\efg{\end{figure}}

\begin{document}
\title{Detection of quantum phase boundary at finite temperatures in integrable spin models}

\author{Protyush Nandi} 

\email{protyush18@gmail.com}

\affiliation{Department of Physics, University of Calcutta,
92 Acharya Prafulla Chandra Road, Kolkata 700009,
India}

\author{Sirshendu Bhattacharyya}

\email{sirs.bh@gmail.com}

\affiliation{Department of Physics, Raja Rammohun Roy Mahavidyalaya,
Radhanagar, Hooghly 712406, India}

\author{Subinay Dasgupta}  

\email{sdphy@caluniv.ac.in}

\affiliation{Department of Physics, University of Calcutta,
92 Acharya Prafulla Chandra Road, Kolkata 700009,
India}

\begin{abstract}
\noindent
Quantum phase transitions occur when quantum fluctuation destroys order at zero temperature. With an increase in temperature, normally the thermal fluctuation wipes out any signs of this transition. Here we identify a physical quantity that shows non-analytic behaviour at finite temperatures, when an interaction parameter is quenched across the line of quantum phase transition. This quantity under consideration is the long time limit of a form of quantum fidelity. Our treatment is analytic for XY chain and 2D Kitaev model and is numerical for a 3D Hamiltonian applicable to Weyl semimetals.\\ 
\end{abstract}
\maketitle

\noindent
The dynamics of quantum many-body system at non-zero temperatures has always been an intriguing area of study, primarily because of the interplay between the quantum and the thermal fluctuations \citep{hofferberth2008,campisi2011,sachdev_2011}. The dominance of thermal fluctuation with increasing temperature makes the perception of quantum noise limited to low temperatures only \citep{greiner2002,haller2010,zhang2012,guan2013}. The question is, whether a quantum phase transition (QPT), exclusively driven by quantum fluctuations at zero temperature, has any impact on the behaviour of the system at non-zero temperature and whether some physical quantity measured at finite temperature bears the signature of the QPT occurring at zero temperature \citep{sondhi1997}. Over the past decades, this issue has been addressed through the studies of quantum fidelity. At zero temperature, fidelity generally vanishes in the thermodynamic limit at a quantum critical point as on two sides of this point the ground state wave functions are structurally different (Anderson's orthogonality catastrophe) \citep{zanardi2006,zanardi2007jsm,cozzini2007,zhou2008,gu2010,damski2016}. At finite temperatures, generalized forms of fidelity have been studied in different systems \citep{jacobson2011,zanardi2007,amin2018,quan2009,dai2017,liang2019,michal2021}, and some of them do detect the QPT through non-analytic signature in their logarithms 
{ at low 
temperatures  \citep{jacobson2011,zanardi2007}. An important work in this direction is by Li, Zhang and Lin \cite{Li_new} who have calculated quantum coherence for XXZ chain using transfer matrix renormalisation group technique and could detect the presence of QPT at finite values of temperature. 
Recently, Hou et. al. \cite{Hou2022} have studied at finite temperatures, a form of rate function for Loschmidt amplitude and detected the presence of QPT.
However, all these works were limited to 1D systems and were not applicable to very high temperatures. } 

The objective of this paper is to look for a quantity that has a robust non-analytic behaviour at zero 
as well as finite temperature while moving across the quantum phase boundary through the quench of a parameter. 
At a temperature $T$, we perform a sudden quench of the Hamiltonian from $\mathcal{H}$ to $\mathcal{H}^{\prime}$ at time $t=0$ and define quantum fidelity as
\be \mathcal{F}_t \equiv \frac{{\rm Tr} \left[ \rho_t \cdot \rho_0 \right]}  {{\rm Tr} \left[ \rho_t \right]\, {\rm Tr} \left[ \rho_0 \right]} \label{def-Fd} \ee
where $\rho_0$ is the density matrix at $t=0$ and $\rho_t$ is the same after the system has evolved for time $t$ under the Hamiltonian $\mathcal{H}^{\prime}$. At zero temperature, this expression reduces to the usual expression for the probability, $|\langle \psi(0) |\psi(t) \rangle|^2$ (where $|\psi(t)\rangle$ is the normalized wave function at time $t$) called the Loschmidt echo and the logarithm of it shows singularities as a function of time, indicating a dynamical quantum phase transition \citep{Heyl_2018}. However, at finite temperatures, there is no such singularity {(See however \cite{Hou2022}).}
Since logarithm of $\mathcal{F}_t$ is proportional to the system size, we may define a measurable quantity called rate function as,
\be r(t,\beta) \equiv  - \lim_{N\to\infty}\frac{1}{N} \log {\mathcal F}_t \label{r} \ee
{The quantity which turns out to be useful is the {\em long-time average of the rate function}, defined as
\be r_a \equiv \lim_{\tau \to \infty} \frac{1}{\tau} \int_0^\tau  r(t,\beta) \, dt \label{r-inf}\ee
We shall consider two integrable quantum spin models, namely the XY chain and the Kitaev model on a honeycomb lattice.  
Each of these Hamiltonians show a QCP at $T=0$.
We shall show analytically that the quantity $r_a$ shows a non-analytic behaviour at the QCP at {\em any} finite temperature just as at $T=0$. Of course, there does not exist an actual quantum phase transition at $T>0$ but our detector bears a signature of the zero-temperature QCP even at $T>0$. One important strength of our detector is that for a $d$-dimensional lattice, the calculation of the relevant quantity boils down to the evaluation of a $d$-dimensional integral. This enables our method to be applicable to higher dimensional systems. In fact, we shall also show (numerically) that for a three-dimensional Hamiltonian applicable to Weyl semimetals, the quantity $r_a$ shows a non-analyticity at the phase boundary, although only for low temperatures. We shall now discuss the case of 2D Kitaev model in detail and then go over to XY chain and the Hamiltonian for semimetals.
}

\begin{figure}
\includegraphics[scale=0.2]{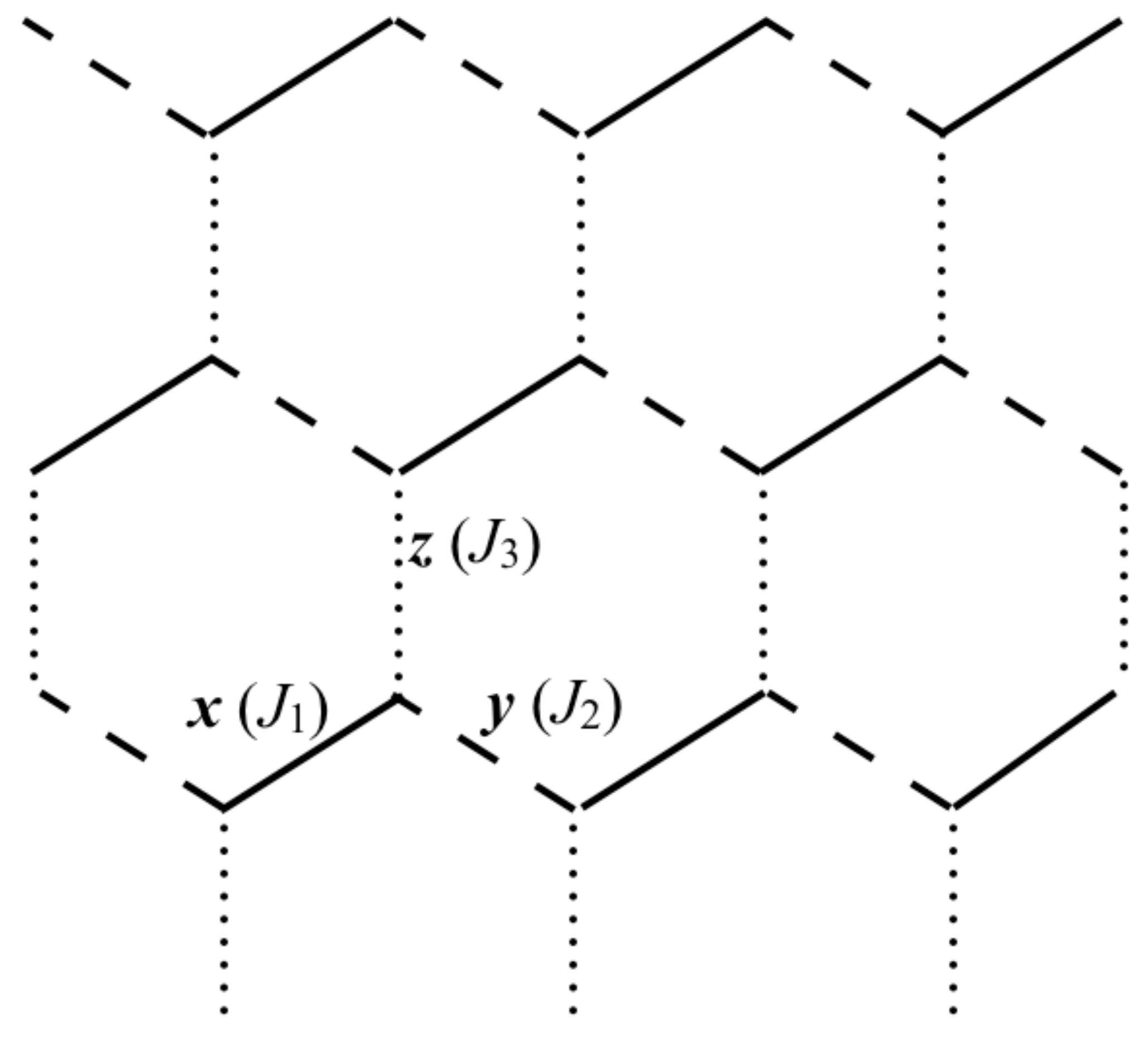}
\includegraphics[scale=0.7]{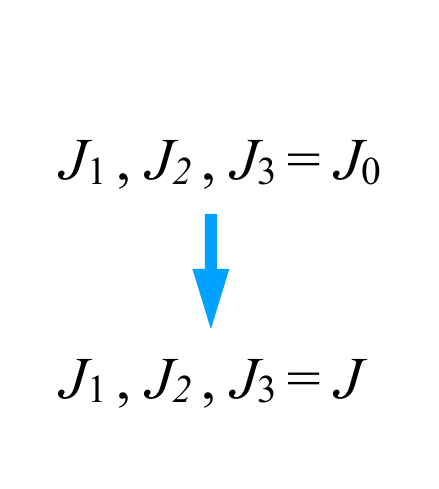}
\caption{Kitaev model on honeycomb lattice. The continuous, dashed and dotted lines correspond to $xx$, $yy$, and $zz$ interactions respectively. We study a quench, where $J_1$ and $J_2$ are kept unchanged and $J_3$ is changed instantaneously from $J_0$ to $J$. For most of the results here, we shall keep $J_0=1$.}
\label{Honeycomb}
\end{figure}

The Hamiltonian of the Kitaev spin-$1/2$ model on a honeycomb lattice is defined as ,
\be \mathcal{H} = \sum_{i,j} J_{\alpha} s_i^{\alpha} s_j^{\alpha} \label{H_def1} \ee
where $i$, $j$ run over all the nearest-neighbouring pairs on the honeycomb lattice, $\alpha$ is 1 or 2 or 3 depending on the location of the sites, as shown in Fig. \ref{Honeycomb}, and $s^{\alpha}=\sigma_{\alpha}$ where $\sigma$ are the Pauli spin matrices. This model contains three interaction parameters $J_{\alpha}$. It can be shown that in the vortex-free sector, this Hamiltonian can be written as a sum of commuting Hamiltonians \citep{kitaev2006,feng2007,kitaev2010}
\be \mathcal{H} = \sum_{\vec{q}} \mathcal{H}_{\vec{q}}, \;\;\;\;\;  {\mathcal H}_{\vec{q}} = a_{\vec{q}} \sigma_3 + b_{\vec{q}} \sigma_1  \label{H_def2}      \ee
where each component of ${\vec{q}}=(q_x, q_y)$ spans over {a square lattice in the range }$(-\pi,\pi)$ and the coefficients are given by \citep{sengupta2008,senguptaPRB2008}
\bea a_{\vec{q}} &=& J_3 -  J_1 \cos q_x -  J_2 \cos q_y, \nonumber \\
b_{\vec{q}} &=& - J_1 \sin q_x +  J_2 \sin q_y \label{ak-bk}\eea
{The ground state shows a gapless phase} in the region satisfying triangular inequality
$|J_i - J_j|  \le  J_k \le J_i + J_j $
where $i$, $j$, $k$ are cyclic permutations of 1, 2, 3. {The gapless and gapped phases are separated by a phase transition line. The two phases are topologically different \citep{kitaev2006,feng2007,kitaev2010} and can be detected by studying Loschmidt echo \cite{Kehrein}. We set $J_1=J_2=1$ so that the critical line occur at $J_3=2$. We shall prove analytically that for a quench $J_3=J_0 \to J_3=J'$, the second derivative (with respect to $J'$) of the rate function $r_a$ diverges algebraically with an exponent $1/2$ at the phase boundary $J'=2$. We shall consider only the vortex-free sector and comment on this aspect later.

} 

The decomposition of the Hamiltonian  in  the Eq~(\ref{H_def2}) enables one to express the density matrix $\rho=\exp (-\beta \mathcal{H})/{\rm Tr}  [\exp (-\beta \mathcal{H})]$  in terms of $\exp \left(-\beta \mathcal{H}_{\vec{q}}\right)$ where $\beta$ is the inverse temperature scaled by Boltzmann constant. To calculate the exponential of $\mathcal{H}_{\vec{q}}$, we write,
\be  {\mathcal H}_{\vec{q}}  = \lambda_{\vec{q}}  {\mathcal G}_{\vec{q}}      \ee
and exploiting the fact $\mathcal{G}_{\vec{q}}^2$ is unit matrix, obtain the quantum fidelity of Eq.(\ref{def-Fd}) and from there the rate function, as
\be r(t,\beta)=  \log 2 - \frac{1}{4\pi^2}\int_{\vec{q}} A_{\vec{q}} \, d\vec{q}      \label{expr-r}  \ee
with 
{ \be A_{\vec{q}} = \log \left[1 + \tanh^2 \beta \lambda_{\vec{q}} \left\{1 - 2\sin^2 \lambda'_{\vec{q}}  t \, \sin^2(\phi_{\vec{q}}  )\right\}\right] \label{def-fq} \ee
where prime refers to the post-quench Hamiltonian and 
$\phi_{\vec{q}} = \theta_{\vec{q}} - \theta'_{\vec{q}}$, where $\theta_{\vec{q}}$ is defined by
\be \cos \theta_{\vec{q}}  = a_{\vec{q}} /\lambda_{\vec{q}} ,\;\;  \sin \theta_{\vec{q}}  = b_{\vec{q}} /\lambda_{\vec{q}} \label{def_theta} \ee
and $\theta'_{\vec{q}}$ by similar expressions with primed quantities.} 

As mentioned earlier, 
at zero temperature, the $\tanh$ term is 1, and $A_{\vec{q}}$ shows singularities as a function of time, but at finite temperature no such singularity occurs since the argument of the logarithm never vanishes. The long-time average of the rate function, as defined in Eq. (\ref{r-inf}) can be calculated from Eq. (\ref{expr-r}) 
using standard results \citep{gradshteyn2014}. 
\bea &&  r_a  =  3\log 2 - \frac{1}{4\pi^2} \int_{\vec{q}} d\vec{q} \; \log(1+\alpha_{\vec{q}})  \nonumber \\
&& - \frac{1}{2\pi^2} \int_{\vec{q}} d\vec{q} \;   \log \left[ 1 + \sqrt{1-\gamma_{\vec{q}} \sin^2(\theta_{\vec{q}} - \theta'_{\vec{q}})}\right] \label{r_inf2} \eea
where $\alpha_{\vec{q}} = \tanh^2 \beta \lambda_{\vec{q}}$, $\gamma_{\vec{q}}=2\alpha_{\vec{q}}/(1+\alpha_{\vec{q}})$.
 
A few subtle issues need to be discussed. (1) The angle $\theta$ ($\theta'$) is undefined where $\lambda$ ($\lambda'$) is zero in the $\vec{q}$ plane, but this fact will not spoil the integration in Eq.~(\ref{expr-r}) because we may exclude small regions ${\mathcal R}$ and ${\mathcal R}'$ around $\lambda=0$ and $\lambda'=0$ respectively, from the integral and evaluate it in the limit ${\mathcal R} \to 0$, ${\mathcal R}' \to 0$. (2) In Eq. (\ref{r-inf}), the quantity $\lambda'\tau \to \infty$ as $\tau \to \infty$ since the point $\lambda'=0$ is excluded from the integration. (3) The long time limit of fidelity has also been studied in Ref.~\citep{zhou2019}, but our procedure of calculating the long-time limit is different from theirs. They have first calculated the long-time limit of fidelity and then taken the logarithm to get the rate function while we have taken the long-time limit of the rate function itself, since the experimentally measurable quantity is the rate function \citep{jurcevic-LE-expt} and not the fidelity  $ \mathcal{F}_t$. (4) When $\rho_t$ in Eq.~(\ref{def-Fd}) is replaced by the {\em equilibrium} density matrix for the post-quench Hamiltonian, namely, 
$\rho_0({\mathcal H}') = \exp\left(-\beta \mathcal{H}'\right)$  
one gets a measure of fidelity \citep{wang2008,bo2012} different from ours, since the Hamiltonian being integrable, the $t \to \infty$ limit of $\rho_t$ is not the same as $\rho_0({\mathcal H}')$. 

In this work, we shall study a particular type of quench where the interaction parameters $J_1$ and $J_2$ of the Kitaev Hamiltonian Eq.~(\ref{H_def1}) is kept fixed at 1, so that the gapless phase exists for $0\leq J_3 \leq 2$ and the gapped phase, for $J_3 > 2$. We quench the parameter $J_3$ from $J_0$ to $J$ and study how the rate function $r_a$ depends on $J$ when $J$ approaches 2 from being within the gapless phase.
 
Thus, the first and second derivatives of the rate function is obtained as (omitting the suffix $\vec{q}$)
\bea \frac{\partial r_a}{\partial J} & = & - \frac{1}{4\pi^2} \int_{\vec{q}} \frac{d\vec{q}}{\lambda'} \; BC \label{d_r}\\
\frac{\partial^2 r_a}{\partial J^2} & = &  - \frac{1}{4\pi^2} \int_{\vec{q}} \frac{d\vec{q}}{\lambda'^2} \; \left[ - \left(1+  \frac{1}{2D}\right) B^2 C^2 \right. \nonumber \\
&& \left. - 2\alpha B \sin \theta' \sin(3\theta' - 2\theta) \right] \label{dd_r}\eea
with $B=(2-\gamma)/(D +D^2)$, $C = \alpha \sin \theta' \sin(2\theta' - 2\theta)$  and $D=\surd[1-\gamma \sin^2 (\theta' - \theta)]$.

\begin{figure}
\includegraphics[scale=0.23]{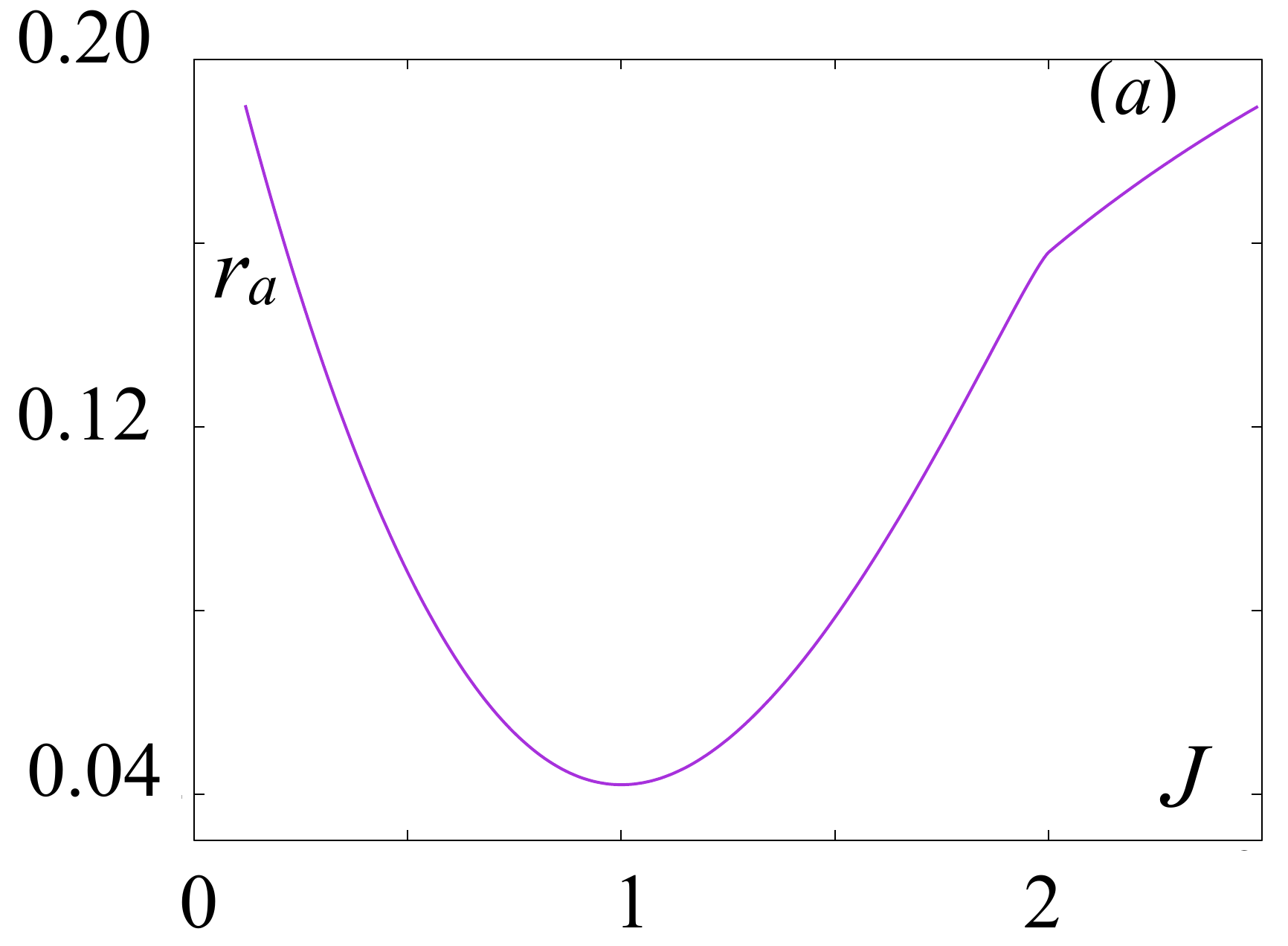}
\includegraphics[scale=0.24]{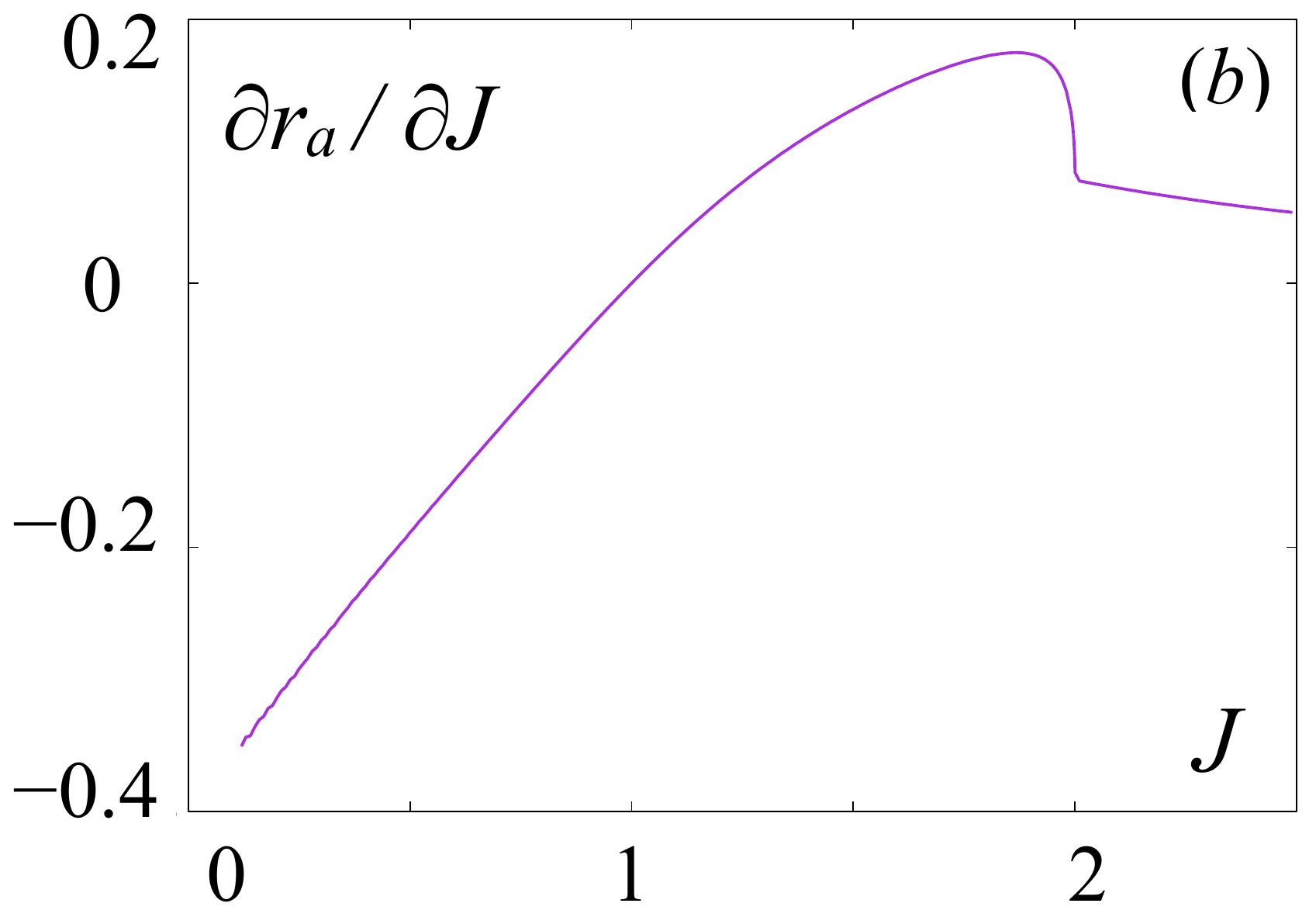}
\includegraphics[scale=0.24]{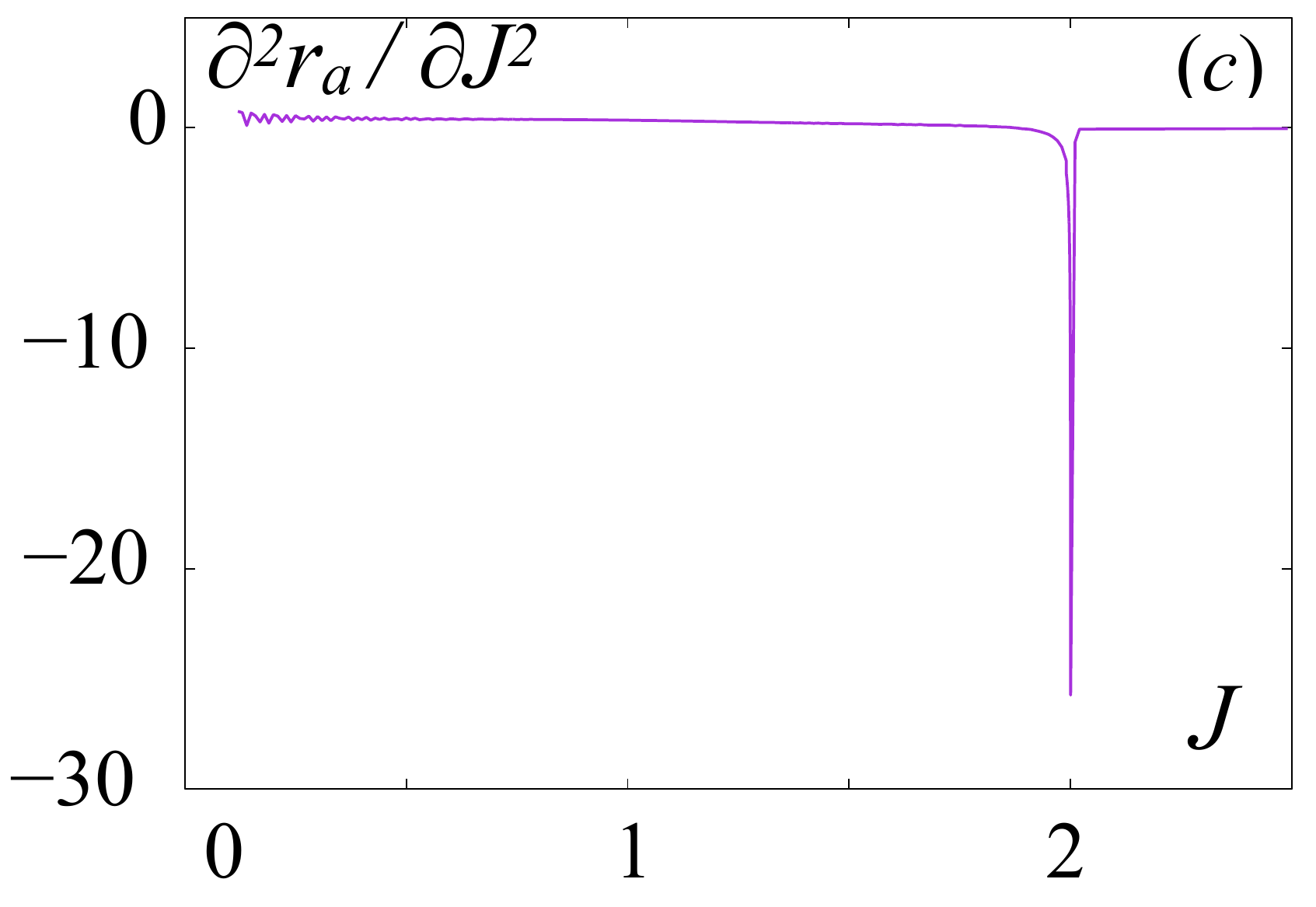}
\includegraphics[scale=0.23]{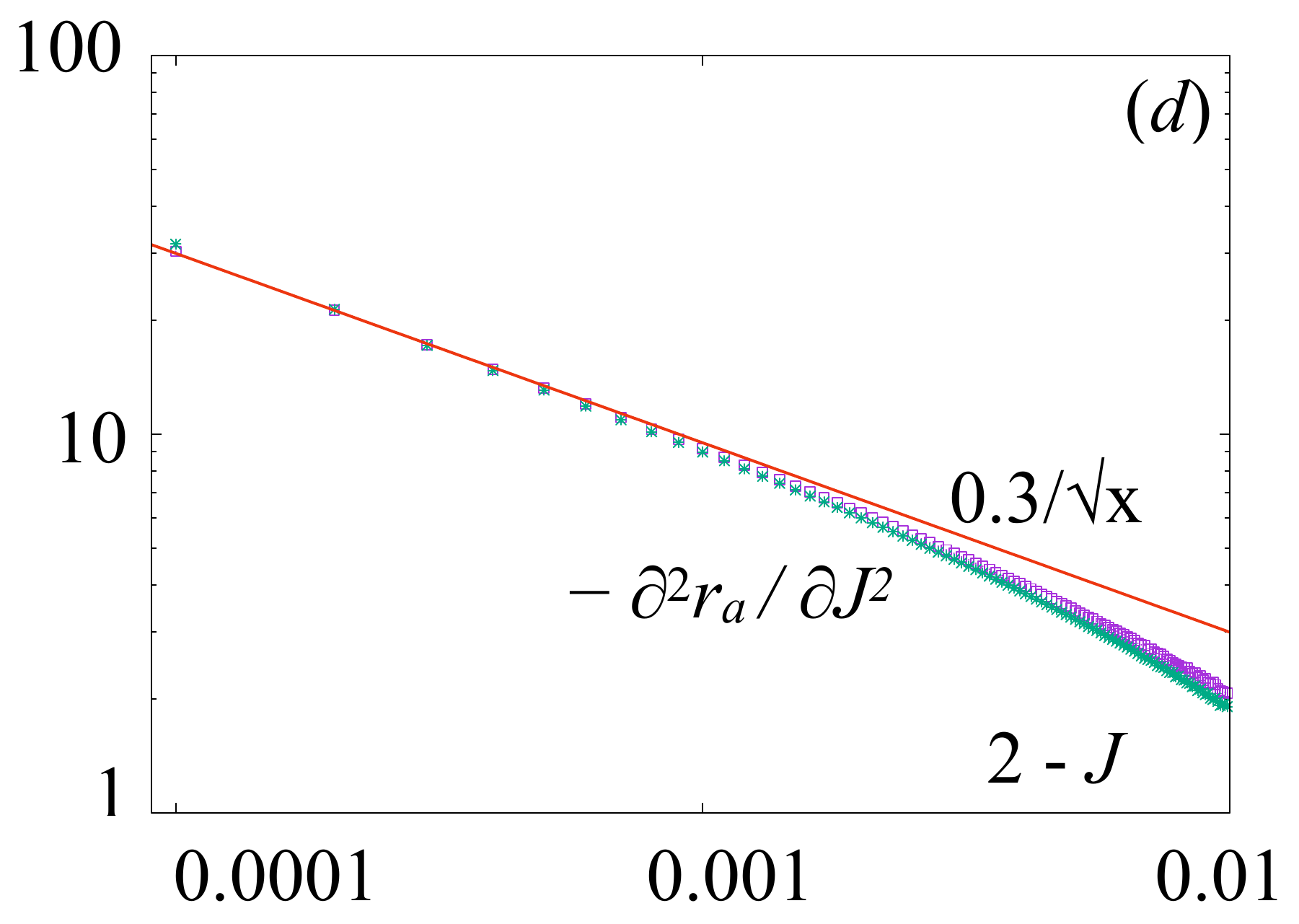}
\caption{Rate function $r_a$ and its derivatives computed numerically from Eqs. (\ref{r_inf2},\ref{d_r},\ref{dd_r}) with $J_0=1$ and inverse temperature $\beta=2$. Non-analyticity appears only at the phase boundary $J=2$. Fig.~(d) shows the scaling of the second derivative as $J$ approaches the value 2 from below. Also, in (d) the violet squares are obtained by integrating over the entire region $-\pi < (q_x, q_y) < \pi$ while the green crosses are obtained for $-0.1 \pi < (q_x, q_y) < 0.1 \pi$. This shows that only the region near the origin is important for the nature of divergence.}
\label{K2D_Results}
\end{figure}

Numerical integration shows that when $J$ {approaches the phase boundary from below}, there appears a non-analyticity {\em at any finite temperature} - the rate function shows a kink, the first derivative remains continuous but undergoes a change of slope, while the second derivative shows power-law divergence with exponent $1/2$ {(Fig. \ref{K2D_Results})}. The expressions for the first and second derivatives contain $1/\lambda'$ and $1/\lambda'^2$ respectively in the integrand. Indeed, whenever $J$ lies within the gapless phase, the region of integration includes a point $\vec{q} = \vec{q_c}$ where $\lambda'$ vanishes. However, the presence of this point leads to a non-analyticity only when $J$ is {\em on} the phase boundary (see the Supplemental Material).

We now set out to study analytically the behaviour of the rate function as a function of the post-quench parameter $J$.  Instead of using Eqs.~(\ref{d_r}, \ref{dd_r}), we shall rather start with the expression of $r_a$ as in Eq.~(\ref{r_inf2}). Using the power series expansion of $\log \left( 1+\sqrt{1-x}\right) $ for any $x$ in the range $0<x<1$, we obtain
\bea r_a  & =  & \log 2 -  \frac{1}{4\pi^2}\int_{\vec{q}} d\vec{q} \log(1+\alpha) \nonumber \\
&& + \frac{1}{8\pi^2} \sum_{n=1, 2, \cdots} c_n \int d\vec{q} \;  \gamma^n \sin^{2n} (\theta - \theta')   \label{rate_expr2} \eea
with $c_1=1$, $c_2=\frac{3}{8}$, $c_3=\frac{5}{24}$ etc. We now express the integrand as
\be \gamma^n \sin^{2n}(\theta - \theta')   = \left(\frac{\tanh \beta \lambda}{\lambda}\right)^{2n}  \left( \frac{2(J-J_0)^2}{1+\tanh^2 \beta \lambda}\right)^n  \left(\frac{b'}{\lambda'}\right)^{2n}  \label{expr1} \ee
and observe that any non-analytic behaviour of this function may arise, if at all, only from a small region around the point where $\lambda'=0$. The location of this point is given by $\vec{q}=(q_c, q_c)$ with $q_c=\cos^{-1}(J/2)$ for $J\le2$. Around this point $\lambda \approx |J_0 - J|$ and  {when $J$ is close to 2 from below, we obtain
\bea r_a & = & \log 2 -  \frac{1}{4\pi^2}\int_{\vec{q}} d\vec{q} \log(1+\alpha) \nonumber \\
&& + \frac{1}{8\pi^2} \sum_n c'_n  \int d\vec{q} \;  \left(\frac{(J-J_0) b'}{\lambda'}\right)^{2n}  \label{rate_expr3} \eea
where $c_n'= c_n[2 \tanh^2 \beta(J_0-2)/(1+\tanh^2 \beta(J_0-2))]^n$.
Indeed, this equality will not work away from the phase boundary.} Numerical results also support this equality (Fig. \ref{K2D_Results}).
Hence, we only need to calculate the integral
\be I_n \equiv (J-J_0)^{2n} \int_{q_x, q_y=-\pi}^{\pi}   dq_x\, dq_y \;  \left(\frac{b'}{\lambda'}\right)^{2n}   ,\;\;\; n=1,2, \cdots  \label{I-def-uv} \ee
As we are interested only in the behaviour of rate function when the post-quench parameter $J$ approaches the value 2 within the gapless phase, we introduce a parameter $\epsilon$ by $J = 2-\epsilon^2$, $ \epsilon=\sqrt{2-J}$
and express $I_n$  as a power series:
\be I_n = a_0 + a_1 \epsilon  + a_2 \epsilon^2  + a_3 \epsilon^3  + \cdots \label{In1} \ee
 It can be shown that (see the Supplemental Material) upto leading order in $\epsilon$,
for any value of $n$, $a_1=0$ and $a_3\ne 0$. In view of Eq.~(\ref{rate_expr3}), this proves that at any temperature,
\be \frac{\partial^2 r_a}{\partial J^2} \sim \frac{1}{\sqrt{2-J}} \label{Final1}\ee
{We mention here that a quantity related to fidelity has been previously observed to show logarithmic divergence in zero temperature \citep{zhao2009}. We also mention some related results of interest: (i) If the pre-quench value of $J_0$ is chosen to be in the gapped phase, the divergence with respect to variation of $J$ remains unchanged  (see the Supplementary Material). (ii) No singularity in rate function is observed when it is studied as a function of  $J_0$ (which is not surprising, since the right-hand side of Eq. (\ref{expr1}) does not show any non-analytic behaviour at $\lambda=0$). (iii) If we approach the phase boundary keeping $J>2$, indeed we observe a non-analyticity but the nature of singularity is different from the one for $J<2$. 
(iv) As we have remained within the vortex-free sector of the Kitaev model, a question arises as to whether the excitation of vortices at finite temperatures destroys the singularity in the rate function $r_a$. In this connection, we note that, the expression of our rate function Eq. (\ref{rate_expr3}) has the form of a series, each term of which is an integral with a pre-factor. The pre-factor involves temperature but not the post-quench value of the parameter, while the integral involves pre- and post-quench value of the parameter but {\em not} temperature. Also, it has been shown \cite{Nasu}  that the vortex excitation in 2D Kitaev model being adiabatic with temperature, does not induce any phase transition.Hence, the pre-factor will not show any non-analytic behaviour as a function of temperature, and it is expected that the rate function will not also show any singularity at a finite temperature due to the presence of vortex excitations. However, for  3D Kitaev model the situation is different since the excitation of vortices triggers a thermal phase transition here. This particular model therefore opens up a future direction of study \cite{Udagawa_2021}.

It is interesting to consider the case of zero temperature. The rate function in Eq. (\ref{r_inf2}) is now,
\be r_a  =  2\log 2 - \frac{1}{2\pi^2} \int_{\vec{q}} d\vec{q} \;   \log \left[ 1 + |\cos(\theta_{\vec{q}} - \theta'_{\vec{q}})| \right]  \label{rate_T0} \ee
This rate function shows a kink at the phase  boundary {\em only}, irrespective of whether the pre-quench parameter $J_0$ is in the gapless phase or in the gapped phase 
(Fig. \ref{T0_Results}). It has been shown \cite{Kehrein}  that the Loschmidt probability as a function of time shows peaks wrongly for quenches within the gapless phase. Hence, we conclude that after taking the long time limit the rate function $r_a$ correctly shows a kink only  at the phase boundary.

\begin{figure}
\includegraphics[scale=0.24]{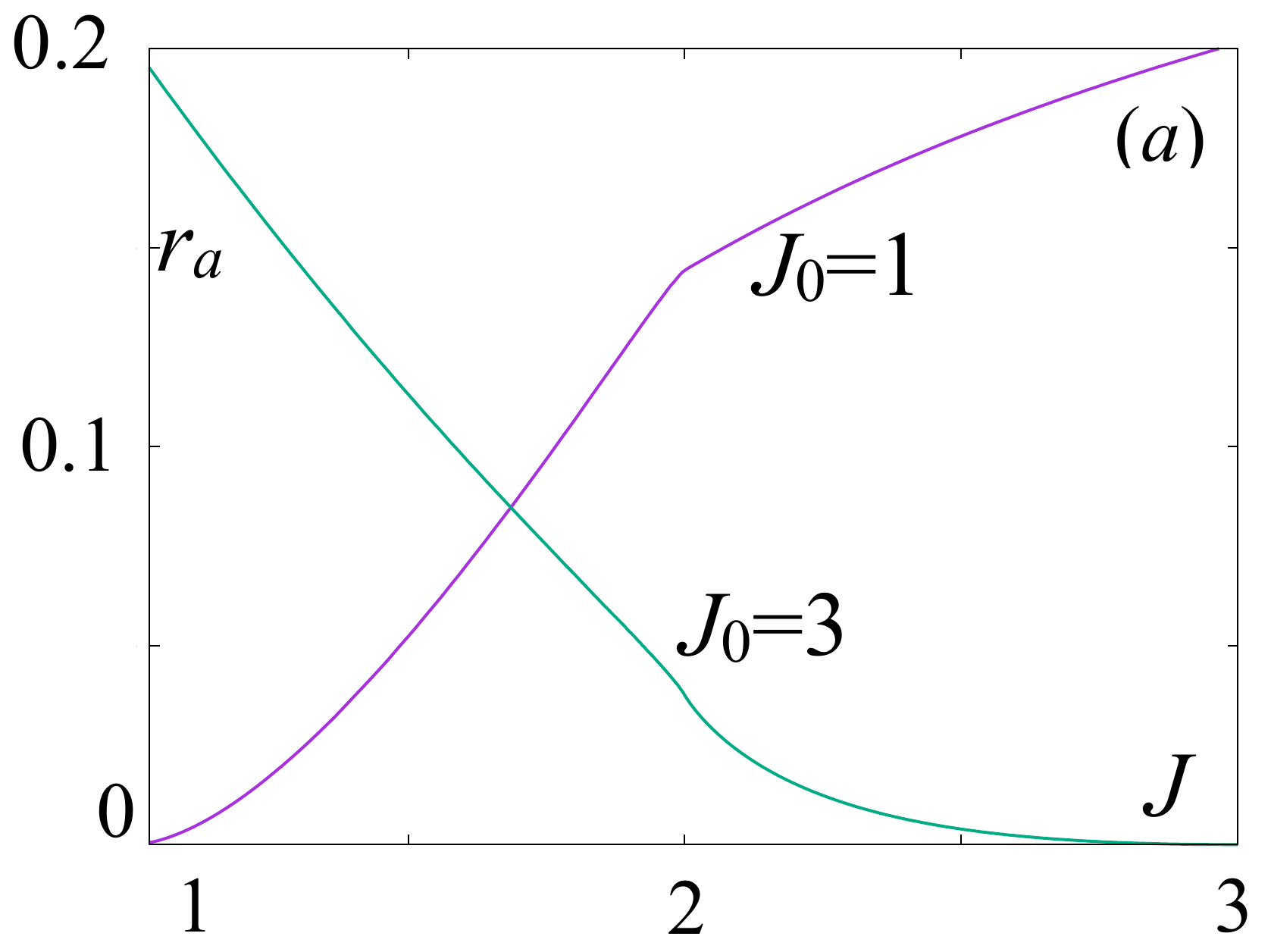}
\includegraphics[scale=0.24]{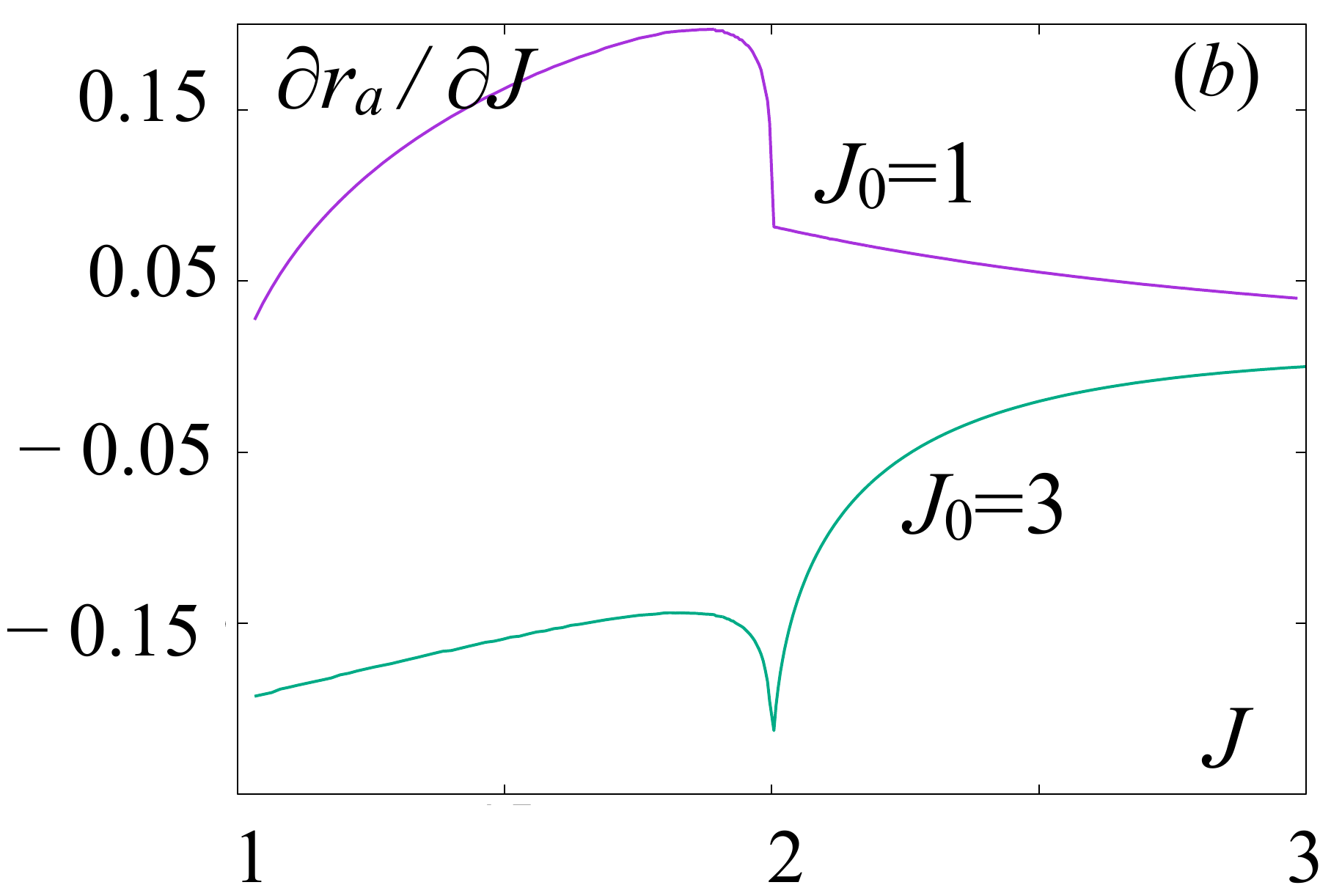}
\caption{Rate function $r_a$ and its derivatives computed from Eq. (\ref{rate_T0}) at zero temperature with the pre-quench parameter $J_0=1$ (in gapless phase) and $J_0=3$ (in gapped phase). Non-analyticity appears only at the phase boundary $J=2$.}
\label{T0_Results}
\end{figure}

We shall now turn to XY Hamiltonian defined by,
\be {\mathcal H}_{XY} = - \frac{1}{2} (1+ h) \sum_{i=1}^N s_i^1 s_{i+1}^1 - \frac{1}{2} (1- h) \sum_{i=1}^N s_i^2 s_{i+1}^2 - \Gamma  \sum_{i=1}^N s_i^3 \label{XY-def1} \ee 
where $h$ is the anisotropy parameter, $\Gamma$ is the transverse field, and $s^{\alpha}=\sigma_{\alpha}$, the Pauli spin matrices. It can be shown by using Jordan-Wigner transformation that this Hamiltonian can be written as a sum of commuting Hamiltonians \citep{sachdev_2011,LSM,*pfeuty} 
\be \mathcal{H} = \sum_q \mathcal{H}_q, \;\;\;\;\;  {\mathcal H}_q = a_q\sigma_3 + b_q \sigma_1  \label{XY_def2}      \ee
where $q$ spans over $(0,\pi)$ and the coefficients are given by 
$ a_q = \Gamma + \cos q$, $b_q = h \sin q \label{XY-ab}$.
The line $-1<\Gamma<1$, $h=0$ separates two ordered phases and the lines $\Gamma=\pm 1$ separate the ordered and the disordered phases. 
Using the expressions for $a_q$ and $b_q$, 
one can define $\theta_q$ from Eq. (\ref{def_theta}) and obtain the rate function and its derivatives from Eqs. (\ref{r_inf2}, \ref{d_r}, \ref{dd_r}) noting that $q$ should now be integrated from 0 to $\pi$. Numerical integration shows that (see the Supplemental Material for figures)
(i) for a quench from $h=h_0 \to h=h'$ (with any $\Gamma$ in the range $-1<\Gamma<1$) the first derivative (with respect to $h'$) of the long-time rate function $r_a$ shows a discontinuity at the QCP $h'=0$ and (ii) for a quench from $\Gamma=\Gamma_0 \to \Gamma=\Gamma'$ at any $h$, the first derivative (with respect to $\Gamma'$) of $r_a$ shows a discontinuity at the QCP $\Gamma'=\pm 1$. Using the approximation in Eq. (\ref{rate_expr3}), one can calculate the  amount of discontinuity at infinitely large temperature (see the Supplemental Material).
\be (\partial r_a /\partial \Gamma')_{\Gamma'=1+} - (\partial r_a /\partial \Gamma')_{\Gamma'=0-} = \beta^2 (1-\Gamma_0^2)/h \ee
\be (\partial r_a /\partial h')_{h'=0+} - (\partial r_a /\partial h')_{h'=0-} = 2\beta^2 h_0^2 (1-\Gamma^2) \ee 

We shall now turn to 3D Hamiltonians. The Hamiltonian for Weyl semimetals with broken time reversal symmetry can be written as \cite{Vishwanath2018, Rao2016},
\be  {\mathcal H}_{\vec{q}}  =  a_{\vec{q}} \sigma_3 +  b_{\vec{q}} \sigma_1 + c_{\vec{q}} \sigma_2 \label{H-WM-def} \ee
where $a_{\vec{q}} =  J_3 - \cos  q_x - \cos q_y -\cos q_z$,  $b_{\vec{q}} = \sin q_x$, $c_{\vec{q}} = \sin q_y$, and $\vec{q}$ runs over a simple cubic lattice in the range $(-\pi,\pi)$. 
The ground state of this Hamiltonian shows a gapless phase for $J_3<3$ and a gapped phase for $J_3>3$. We consider a quench $J_3=J_0 \to J_3=J$ and note that one can arrive at Eqs. (\ref{expr-r}, \ref{def-fq}) in a straightforward manner, with 
$\phi_{\vec{q}}$ being the angle between the unit vectors 
$(b_{\vec{q}}/\lambda_{\vec{q}}, c_{\vec{q}}/\lambda_{\vec{q}}, a_{\vec{q}}/\lambda_{\vec{q}})$ and $(b'_{\vec{q}}/\lambda'_{\vec{q}}, c'_{\vec{q}}/\lambda'_{\vec{q}}, a'_{\vec{q}}/\lambda'_{\vec{q}})$. 
The evaluation of the rate  function  $r_a$ now reduces to the computation of an integral in the $\vec{q}$ space and we observe numerically that the first derivative (with respect to $J$) of $r_a$ shows a change of slope at the QCP $J=3$ both for $J_0 < 3$ and $>3$ (see the Supplementary Material for figures). It is important to mention that, unlike the previous two cases, this singularity is visible only at low temperatures. When the time reversal symmetry is not broken, $c_{\vec{q}} =0$ in Eq. (\ref{H-WM-def}) and one has the topological nodal line semimetals \cite{Okugawa}. In this case also, one observes a change of slope at $J=3$ of the curve $\partial r_a/\partial J$ vs $J$.

\noindent
To conclude, we explore the finite temperature behaviour of three integrable quantum spin models and observe a non-analyticity in the mixed state fidelity at the phase boundary. The rate function can be written (Eq. (\ref{rate_expr3})) as a series, each term of which is an integral independent of temperature with a pre-factor involving temperature. It is the integral from which the non-analytic behaviour originates(at all temperatures in 2D and at low temperatures in 3D). The fact that the quantity in question is insensitive to thermal fluctuations, makes it a potential candidate to be studied experimentally as a good detector of quantum phase transition. It would be interesting to explore how our rate function behaves for other integrable and non-integrable Hamiltonians. Work in this line is under progress.

}

\paragraph*{Acknowledgement:} PN acknowledges UGC for financial support (Ref. No. 191620072523).

\bibliography{dmt}
\bibliographystyle{apsrev4-2}

\end{document}


\title{Supplemental Material: Detection of quantum phase boundary at finite temperatures in two dimensional Kitaev model}

\author{Protyush Nandi} 

\email{protyush18@gmail.com}

\affiliation{Department of Physics, University of Calcutta,
92 Acharya Prafulla Chandra Road, Kolkata 700009,
India}

\author{Sirshendu Bhattacharyya}

\email{sirs.bh@gmail.com}

\affiliation{Department of Physics, Raja Rammohun Roy Mahavidyalaya,
Radhanagar, Hooghly 712406, India}

\author{Subinay Dasgupta}  

\email{sdphy@caluniv.ac.in}

\affiliation{Department of Physics, University of Calcutta,
92 Acharya Prafulla Chandra Road, Kolkata 700009,
India}

\maketitle

\section{\bf Why the quantity $r'' \equiv \partial^2r_a/\partial J^2 $ diverges only at the phase boundary?} 
\begin{figure}[h]
\includegraphics[scale=0.5]{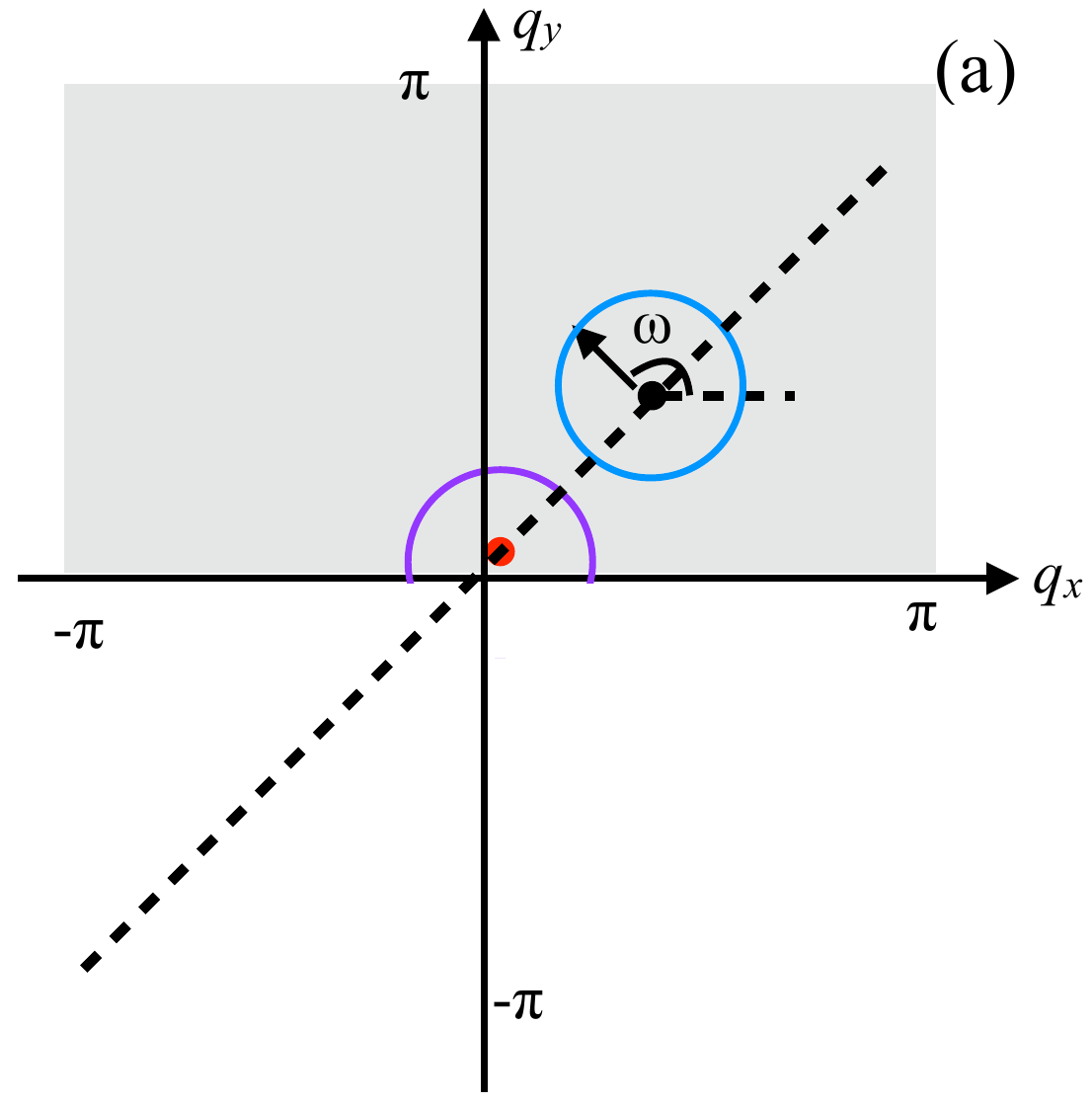}
\includegraphics[scale=0.5]{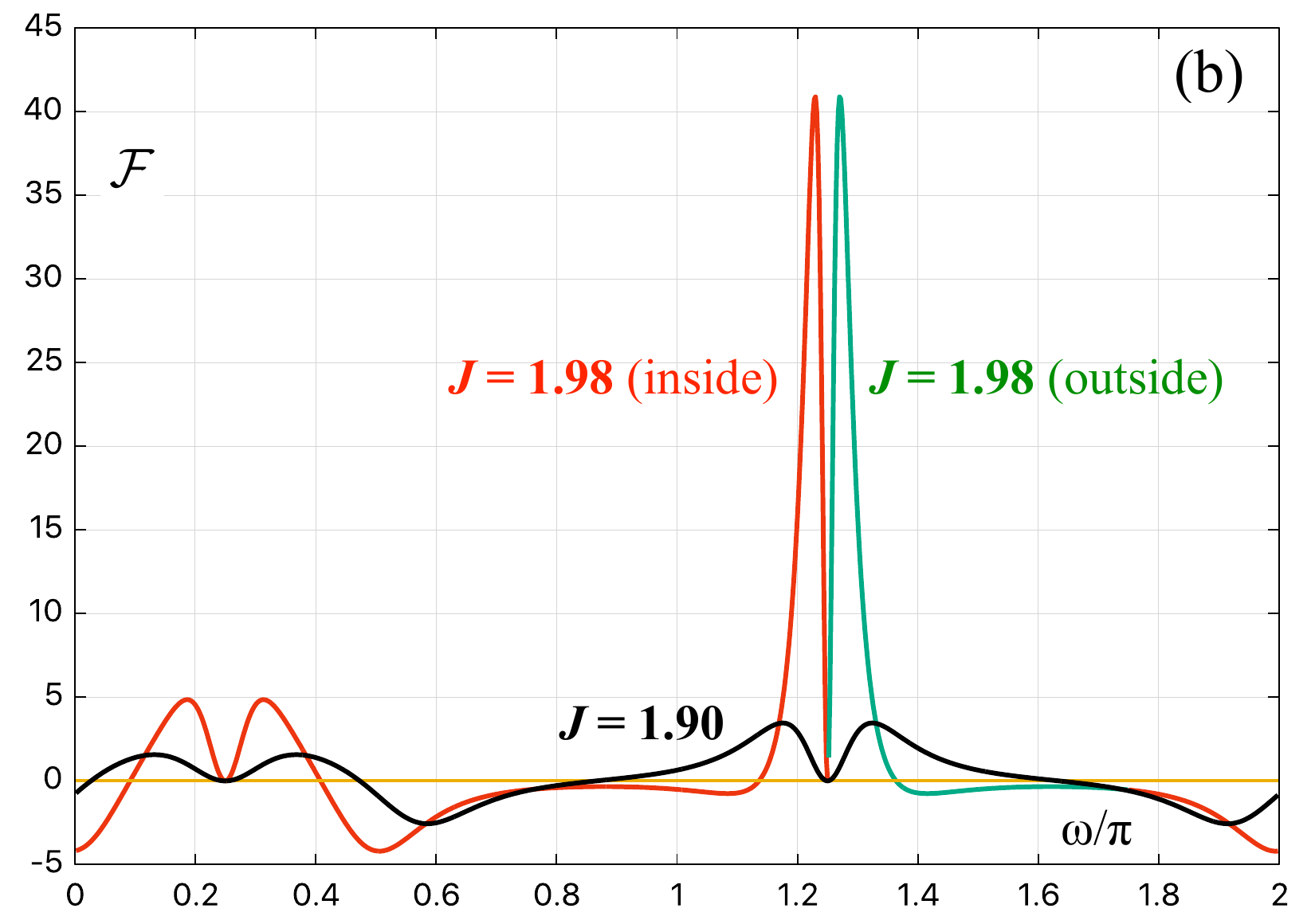}
\caption{
Profile of ${\mathcal F}$, the integrand of $\partial^2r_a/\partial J^2 $ at $\beta=2$. We consider $J=1.90$ and 1.98 for which the node is at $q_{xc}=q_{yc}=0.32$ and 0.14 respectively. (a) Circle around the nodes for two $J$ values. The points on the circle have polar angle $\omega$ measured with respect to the $q_x$ axis. Note that the region of integration runs over positive values of $q_y$.  For $J$ somewhat smaller than $2$, the circle lies entirely within the $q_y > 0$ region while for $J$ close to 2, part of the circle goes outside this region.
(b) Variation of ${\mathcal F}$ over the two circles of radius $R=0.2$. For $J=1.90$, the circle is in the $q_y > 0$ region and the positive and negative values nearly cancel out, while for $J=1.98$ such cancellation does not occur since part of the circumference is excluded. Furthermore, the magnitude of ${\mathcal F}$ is larger in the latter case. }
\label{circles}
\end{figure}

\noindent
The quench under consideration is from $(J_1, J_2, J_3) = (1, 1, J_0)$ to $(1, 1, J)$ and we have from Eq. (\ref{dd_r})
\be r''=\partial^2r_a/\partial J^2 = \int\limits_{q_x, q_y=-\pi}^{\pi} {\mathcal F} dq_x dq_y \;\;\mbox{with} \;\;
{\mathcal F} = \frac{1}{4\pi^2 \lambda'^2} \left[ \left(1+  \frac{1}{2D}\right) B^2 C^2 + 2\alpha B \sin \theta' \sin(3\theta' - 2\theta) \right] \ee
According to Eq. (\ref{def_theta})
\bea \lambda \cos \theta &=& J_0 -  \cos q_x - \cos q_y, \nonumber \\
  \lambda \sin \theta &=& -\sin q_x + \sin q_y, \nonumber \\
   \lambda' \cos \theta' &=& J -  \cos q_x - \cos q_y  \nonumber \\
    \lambda' \sin \theta' &=& -\sin q_x + \sin q_y \label{s1}\eea
We first observe that if $\vec{q}$ is replaced by $-\vec{q}$, the quantities $\theta, \theta'$ becomes $-\theta, - \theta'$. Also, such reversal of sign of $\theta, \theta'$ keeps ${\mathcal F}$ invariant. Thus, it is sufficient to integrate only over the region $-\pi < q_x \le \pi$, $0\le q_y < \pi$ (shaded region in Fig.~\ref{circles}a) and write
\be \partial^2r/\partial J^2 = 2\int\limits_{q_x=-\pi}^{\pi} \, \int\limits_{q_y=0}^{\pi} {\mathcal F} dq_x dq_y \ee

For any $J$ consider a circle of radius $R$ around the point where $\lambda'=0$ (We shall call this point as node). The location of this node is given by $q_{xc}=q_{yc}=\cos^{-1} (J/2)$. The quantity ${\mathcal F}$ diverges at the node and decreases as one moves away from it. When $J$ lies well inside the gapless region, the shaded integration region covers such circles for $R=0$ to a fairly large value. However, when $J$ is close to the phase boundary, the shaded region cannot cover the whole circle unless the radius $R$ is too small (Fig.~\ref{circles}). It is seen numerically (but is difficult to verify analytically), 
that the values of ${\mathcal F}$ plotted over a full circle sums up to zero, although being  asymmetric about zero. Hence after integration, $r''$ vanishes when $J$ is well-inside the gapless region. However, when $J$ is close to the phase boundary, the integration is over part of a circle and the cancellation of ${\mathcal F}$ values does not occur. Hence, $r''$ does not vanish. This effect is aggravated by the fact that ${\mathcal F}$ attains numerically larger values  over a wider region when $J$ is close to the phase boundary, as can be easily seen from Fig.~\ref{circles}.

\section{Evaluation of the integral $I_n$}

\noindent
To evaluate the integral $I_n$ as defined in Eq. (\ref{I-def-uv}), 
let us transform to the variables $u=(q_x + q_y)/2$, $v= (q_y -  q_x)/2$ and set $J_0=1$. This gives, 
\be I_n =  4(J-1)^{2n} \int_{u=0}^{\pi/2} du \; \int_{v=-\pi}^{\pi} dv   \;    \left(\frac{b'}{\lambda'}\right)^{2n}    \label{In-uv} \ee
by using the invariance of the integral under $u \to \pi \pm u$, $v \to \pi \pm v$. Recall that $\theta$ and $\theta'$ are defined by Eq.~(\ref{def_theta}), and the expressions for $a$ and $b$ are now $a =1 -  2\cos u \cos v$, $b = 2 \cos u \sin v=b'$, $a' = J - 2\cos u \cos v$. The crucial step is to express  the integrand as
\be \left(\frac{b'}{\lambda'}\right)^{2n} 
= \left(\frac{\mu}{4J}\right)^n \cdot \frac{  \left(z^2 - 1 \right)^{2n}}{z^n \left(z - \frac{\mu}{J} \right)^n \left(z  - \frac{J}{\mu} \right)^n} \ee
with $z = \exp(iv)$ and $\mu=2\cos u$. This gives
\be I_n =  -\frac{4i(J-1)^{2n}}{(4J)^n} \int_0^{\pi/2} du \,  \oint F^{(n)} dz \label{I1} \ee
where $\oint$ stands for integral over unit circle, and 
\[ F^{(n)} (z) = \frac{ \mu^n \left(z^2 - 1 \right)^{2n}}{z^{n+1} \left(z - \frac{\mu}{J} \right)^n \left(z  - \frac{J}{\mu} \right)^n}  \]
For $u<q_c$, the poles inside the unit circle are $z_1=0$ and $z_2=J/\mu$, while for $u>q_c$, the poles inside are $z_1=0$ and $z_3=\mu/J$. Denoting the residue of $F^{(n)}$ at $z_i$ as $R_i^{(n)}$ for $i=1, 2, 3$,
\be I_n =  \frac{8\pi (J-1)^{2n}}{(4J)^n} \sum\limits_{i=1}^{3}\int_0^{\pi/2} R_i^{(n)} \, du   \label{I2} \ee
We define $\epsilon=\sqrt{2-J}$ so that $q_c=\cos^{-1} (1- \epsilon^2/2 ) = \epsilon + \frac{1}{24} \epsilon^3 + \cdots$ and express $I_n$ as a power series:
\be I_n = a_0 + a_1 \epsilon  + a_2 \epsilon^2  + a_3 \epsilon^3  + \cdots \label{In1} \ee
so that 
\[  \frac{\partial^2 I_n}{\partial J^2} = -\frac{a_1}{4} \cdot \frac{1}{(2-J)^{3/2}} + \frac{3a_3}{4} \cdot \frac{1}{\sqrt{2-J}} + \cdots \]
 It is important to note that the divergence behaviour is controlled by the coefficients $a_1$ and $a_3$ only. 

At high temperatures, only the first term in the summation in Eq.~(\ref{rate_expr2}) is dominant and the residues of the integrand in $I_1$ can be easily calculated as
\be R_1^{(1)}= \frac{J^2 + \mu^2}{J}, \;\;\;\; R_2^{(1)} = \frac{J^2 - \mu^2}{J}, \;\;\;\; R_3^{(1)} = \frac{\mu^2 - J^2}{J} \label{residues} \ee
This gives, upto leading order in $\epsilon$,
\be I_1 =  \pi^2 - \pi^2\epsilon^2 -\frac{8\pi}{3} \epsilon^3 \label{expr_I1} \ee
implying that $\partial^2 I_1/\partial J^2 = -2\pi/\sqrt{2-J} $.

To study the behaviour of $I_n$ for $n>1$, we look at Eq. (\ref{I2}) and observe that the residue at $z_2$ is
\be R_2^{(n)} = \dfrac{1}{(n-1)!}\left[ \frac{\partial^{n-1}}{\partial z^{n-1}} \left\{(z-z_2)^nF^{(n)}\right\} \right]_{z=z_2}  \label{R2n_1}\ee
We define
\[ G(z) = \frac{\mu\left(z^2-1\right)^2}{z\left(z-\frac{\mu}{J}\right)} \;\;\;\;\;\; \mbox{so that}  \;\;\;\;\;\; (z-z_2)^nF^{(n)} = \frac{G^n}{z} \]
and observe that
\be \frac{\partial^{n-1}(G^n/z)}{\partial z^{n-1}} 
= \frac{G}{z}  \left[ n \left(\frac{\partial G}{\partial z} \right)^{n-1} +\; \mathcal{T} \right]
\label{sum} \ee
where $\mathcal{T}$ contains the other terms that will necessarily involve $G^k$ or $G^k/z$ as a factor with $k \ge 1$. 

Obviously, $(G/z)_{z=z_2} = R_2^{(1)}$. Also,
\be (G)_{z=z_2}= -\frac{\epsilon^2}{2}+\frac{u^2}{2}+\frac{\epsilon^2 u^2}{4}-\frac{\epsilon^4}{8}-\frac{u^4}{8}+ \mbox{(higher order terms)}       \;\;\;\; \mbox{and}\;\;\;\left(\frac{\partial G}{\partial z} \right)_{z=z_2} = 2J + \frac{\mu^2}{J} \label{R2n_2} \ee
Hence, upto leading order terms in $\epsilon$,
\[ R_2^{(n)} \approx  \dfrac{1}{(n-1)!}\;n4^{n-1} R_2^{(1)} \]
Using Eqs. (\ref{R2n_2}) and (\ref{sum}) in Eq. (\ref{R2n_1}) and then performing the integration over $u$ in Eq. (21) we can conclude that the second integral in Eq. (\ref{I2}) contributes an $\epsilon^3$ term but no term linear in $\epsilon$.  Similar conclusion will hold for the residue at $z=z_3$ too. This implies that $a_1=0$ and $a_3\ne 0$, not only for $n=1$, but for any $n$. In view of Eqs.~(\ref{rate_expr3}), this completes the proof that at any temperature
\be \frac{\partial^2 r_a}{\partial J^2} \sim \frac{1}{\sqrt{2-J}} \label{Final1}\ee

\newpage
\section{Results for XY chain}

\begin{figure}[h]
\includegraphics[scale=0.3]{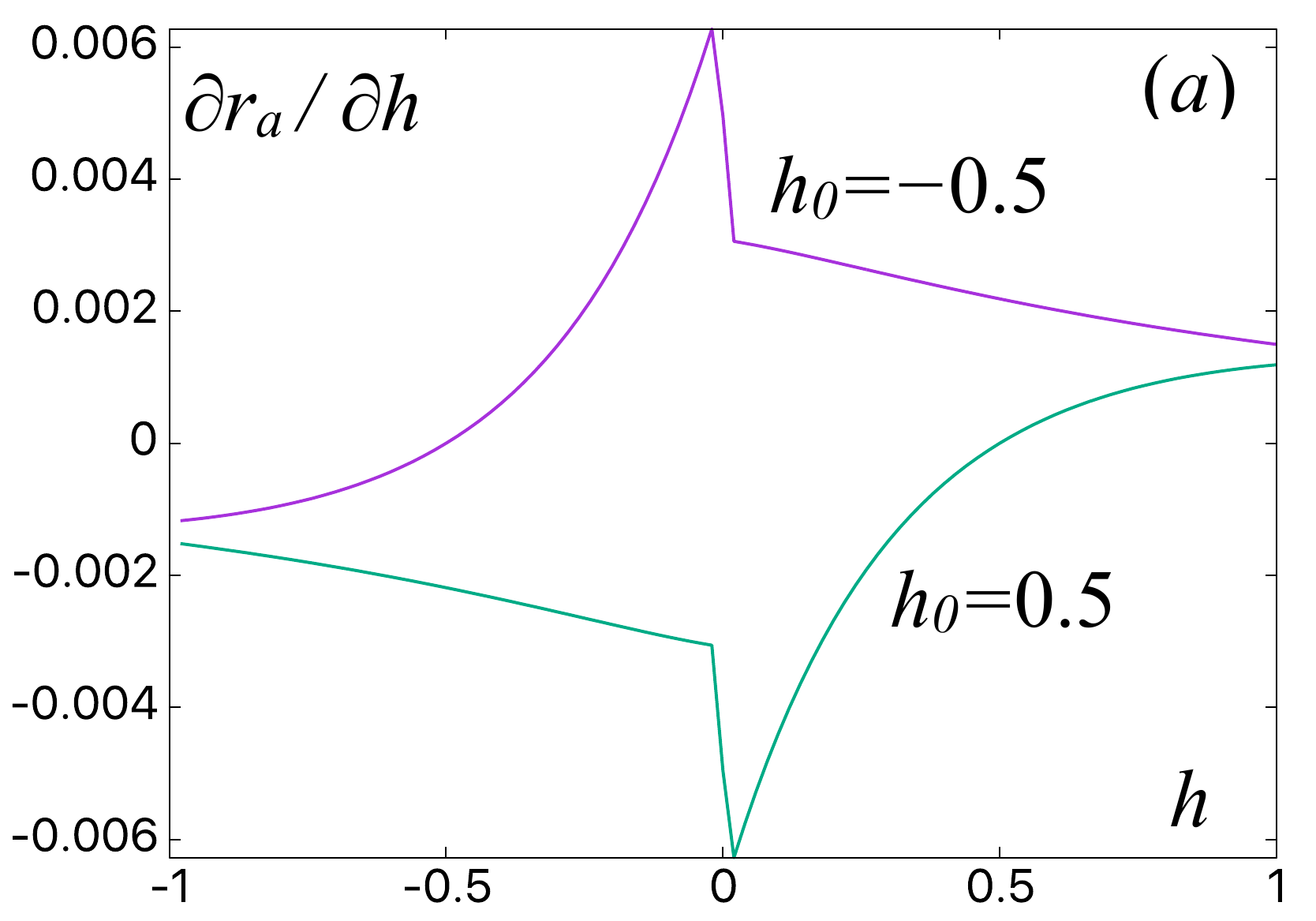}
\includegraphics[scale=0.3]{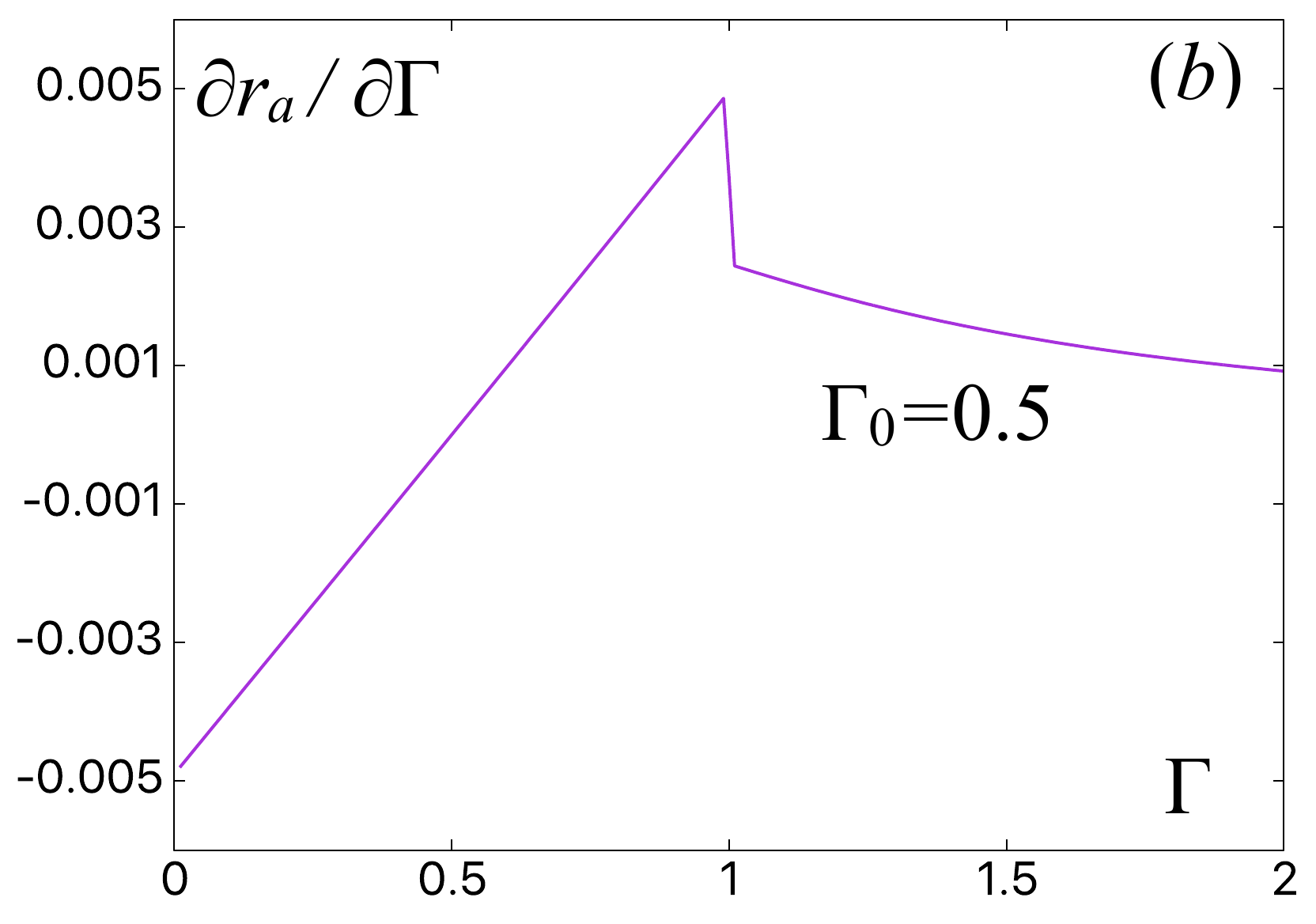}
\includegraphics[scale=0.3]{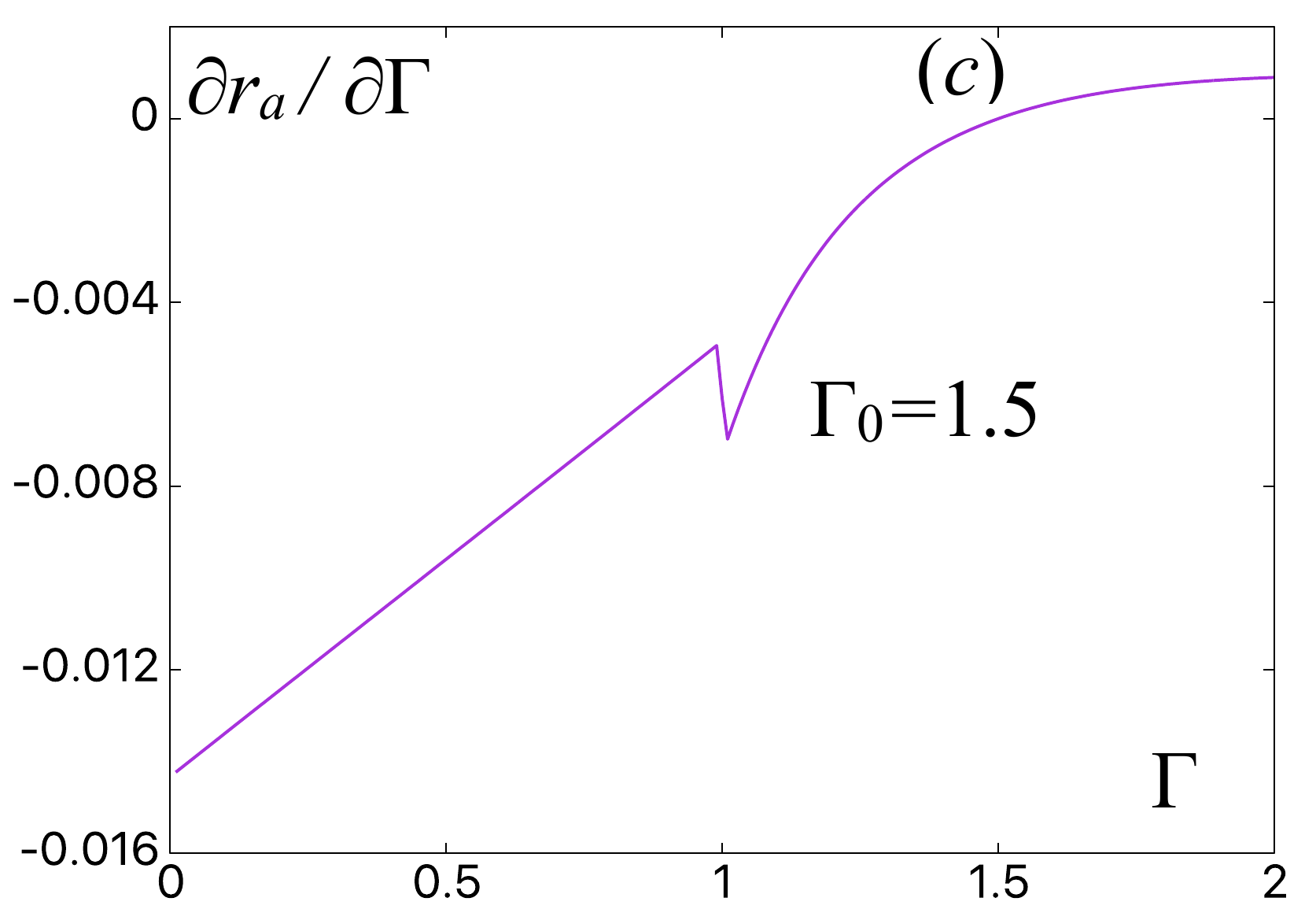}
\caption{ First derivative of rate function computed numerically from Eqs. (\ref{r_inf2}, \ref{d_r}) for quench of the anisotropy parameter at $\beta=0.1$, $\Gamma=0.5$ in (a) and   for quench of the transverse field at $\beta=0.1$, $h=1$ in (b), (c). For (b) $\Gamma_0$ is within the ordered state, while for (c) $\Gamma_0$ is within the disordered state. In both cases the quantity $\partial r_a/\partial \Gamma$ is discontinuous at the critical point $\Gamma=1$. }
\end{figure}

\section{Results for Kitaev model on honeycomb lattice}

\begin{figure}[h]
\includegraphics[scale=0.3]{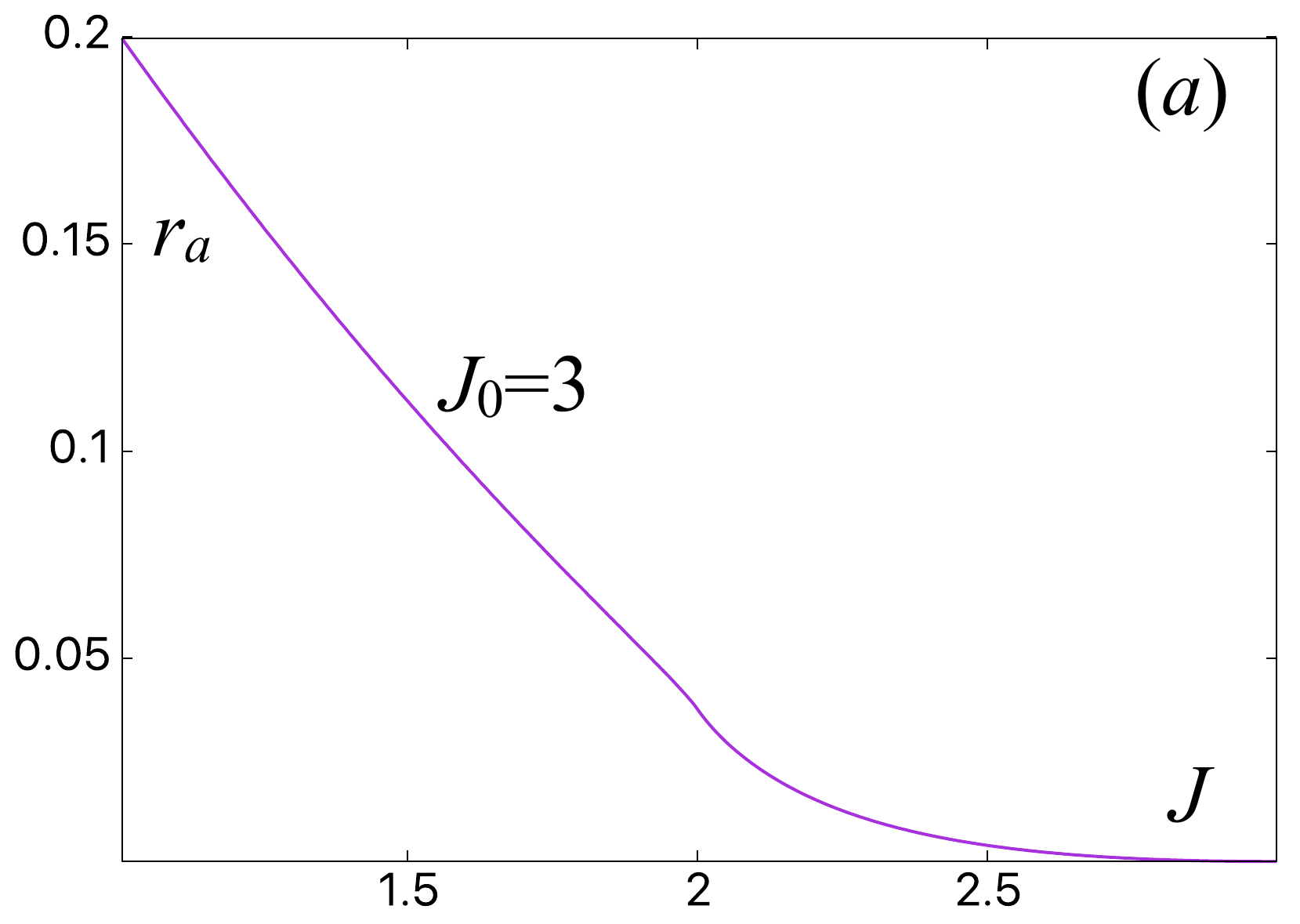}
\includegraphics[scale=0.3]{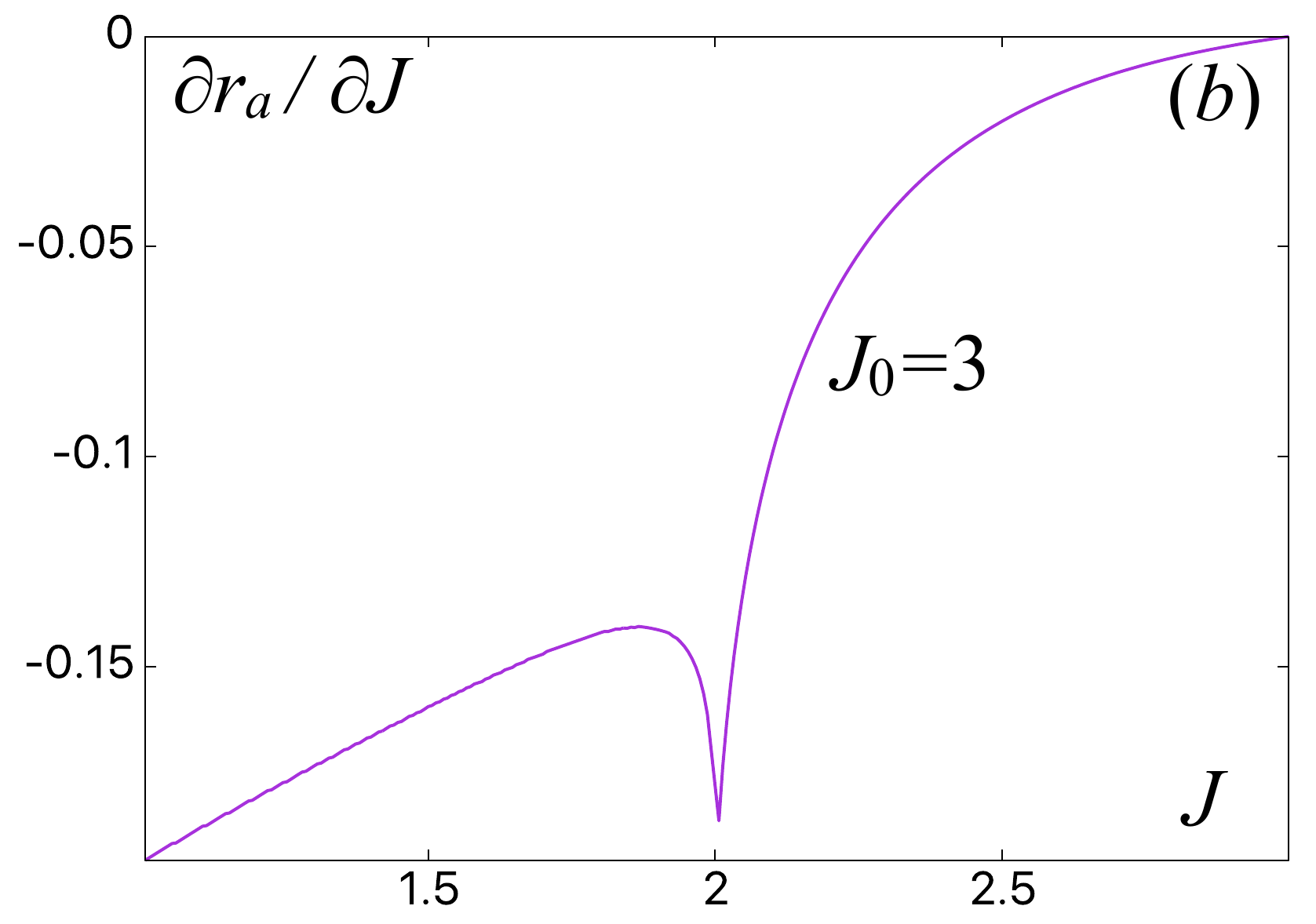}
\includegraphics[scale=0.285]{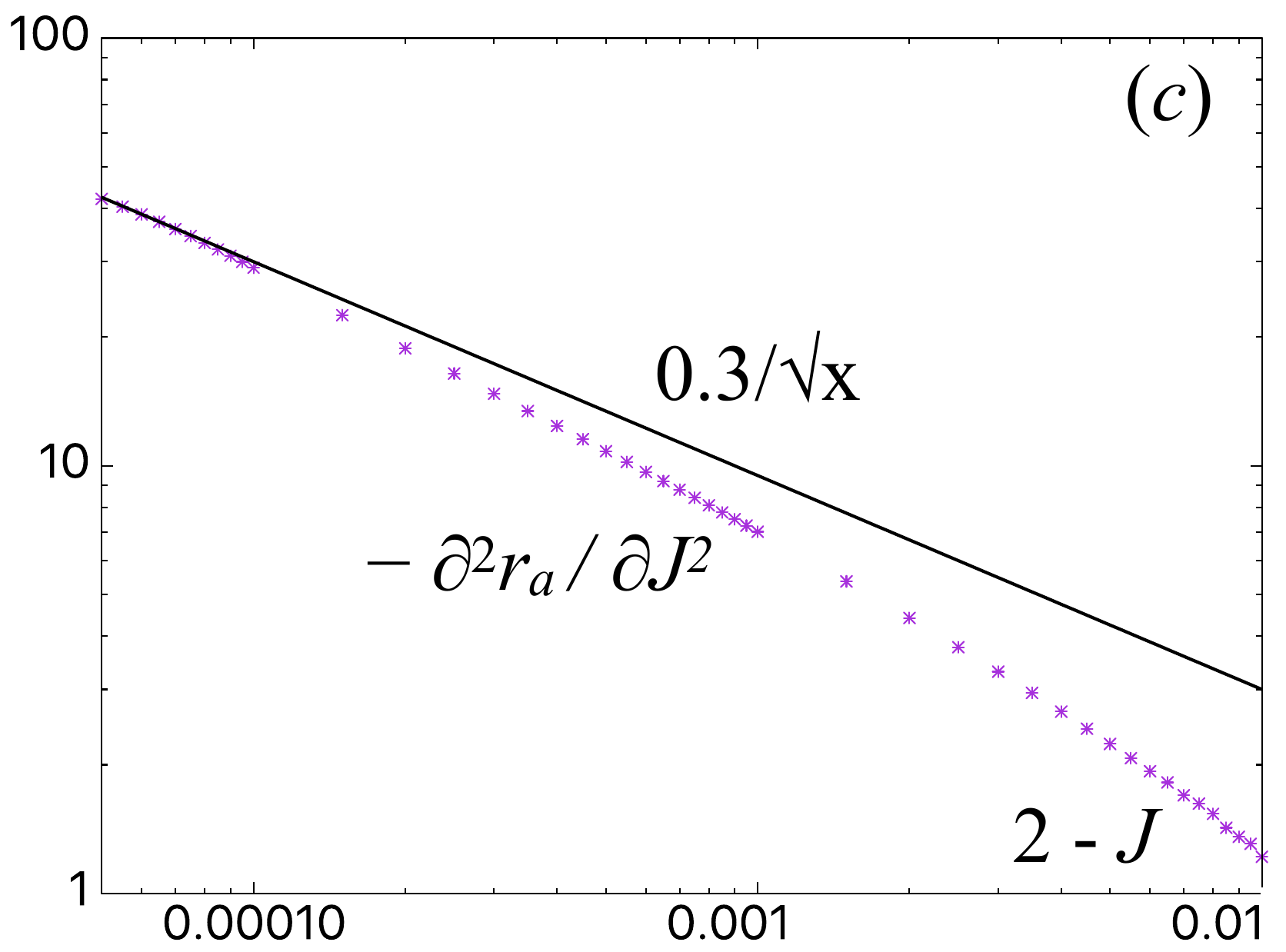}
\caption{Rate function $r_a$ and its two derivatives computed numerically from Eqs. (\ref{r_inf2}, \ref{d_r}) for quench of the parameter $J_3$  at $\beta=2$, $J_1=J_2=1$ with the pre-quench value $J_0=3$. A non-analyticity appears at $J=2$ in the same way as in Fig. 2.}
\end{figure}

\begin{figure}[h]
\includegraphics[scale=0.25]{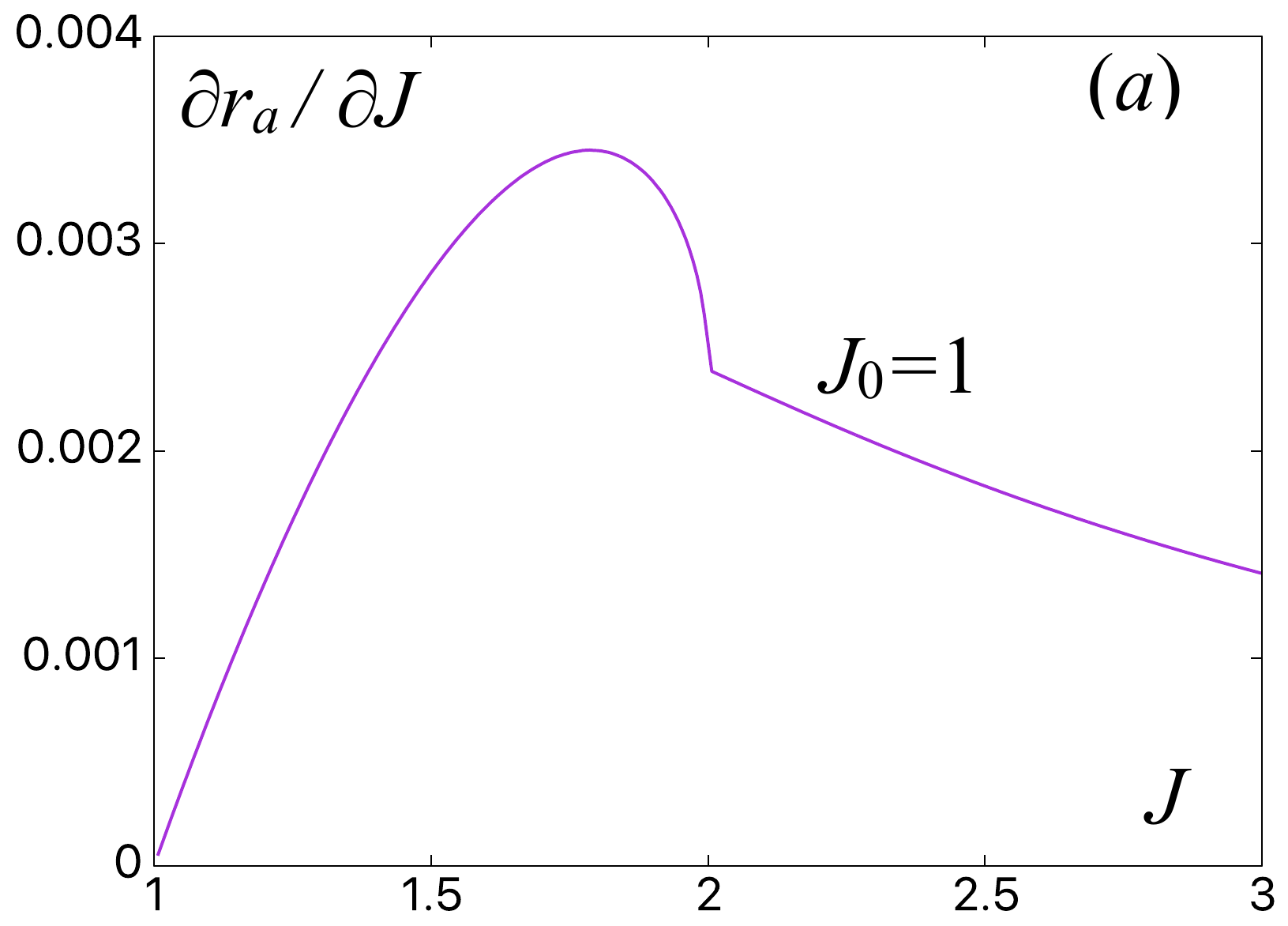}
\includegraphics[scale=0.235]{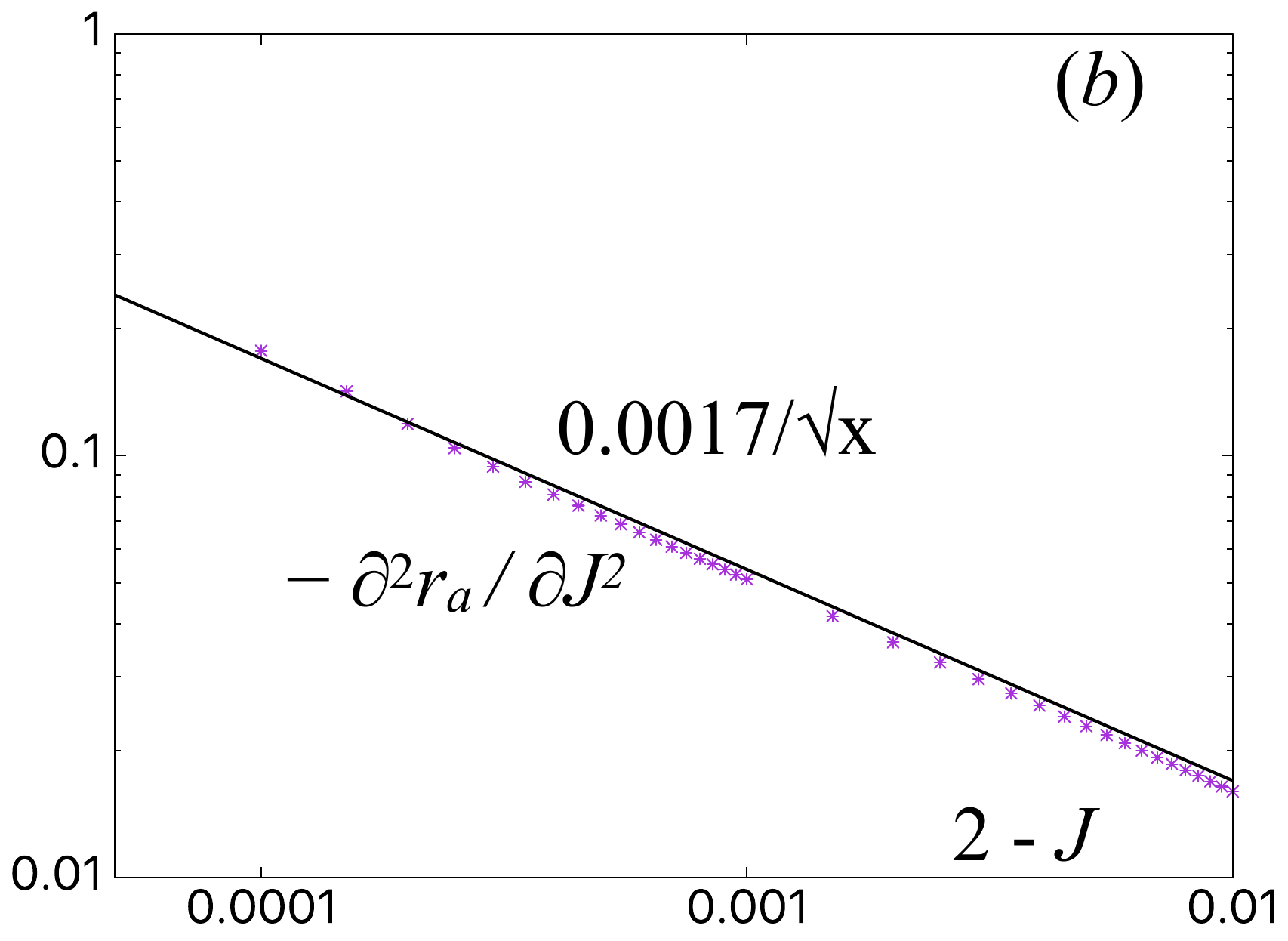}
\includegraphics[scale=0.25]{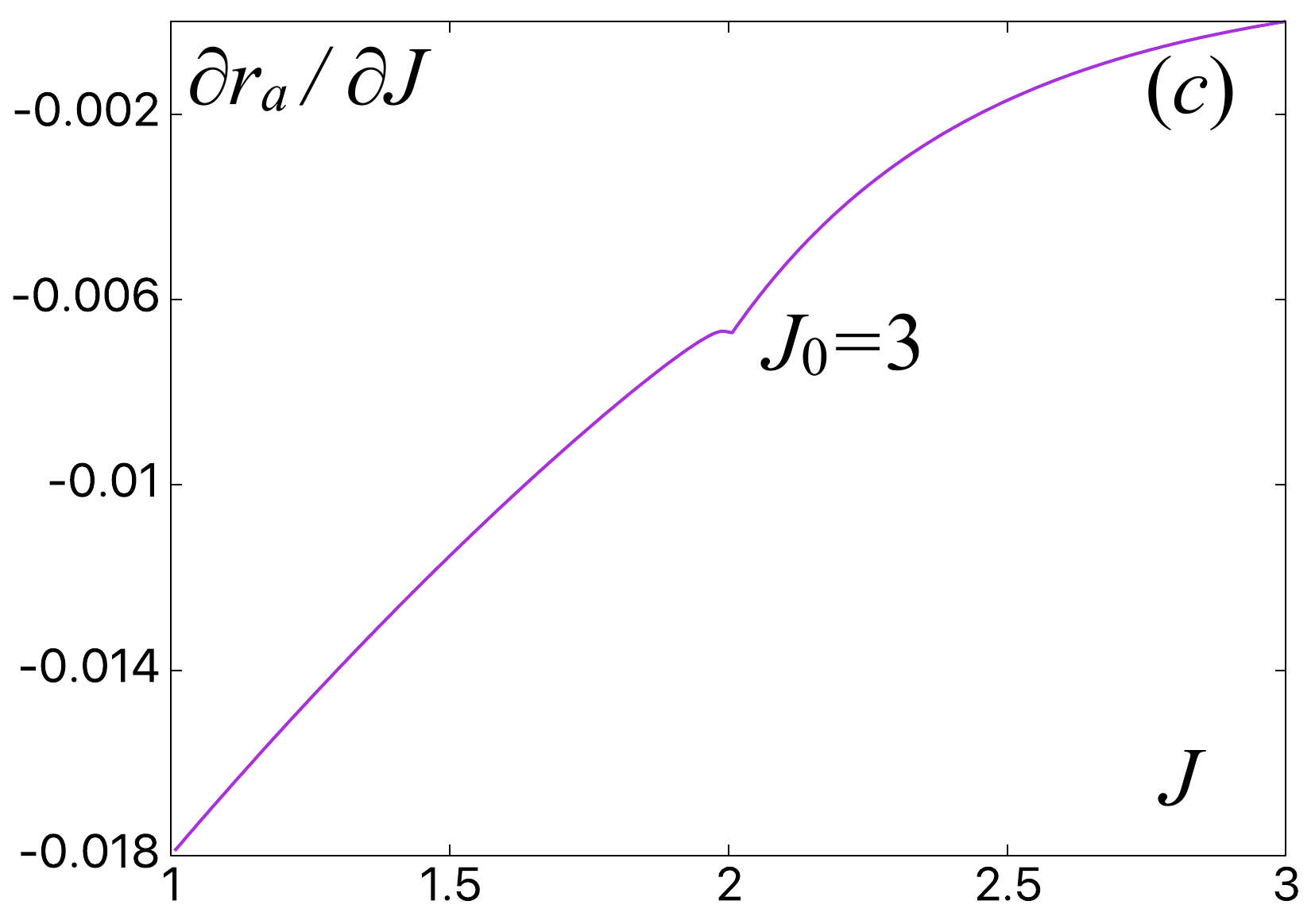}
\includegraphics[scale=0.22]{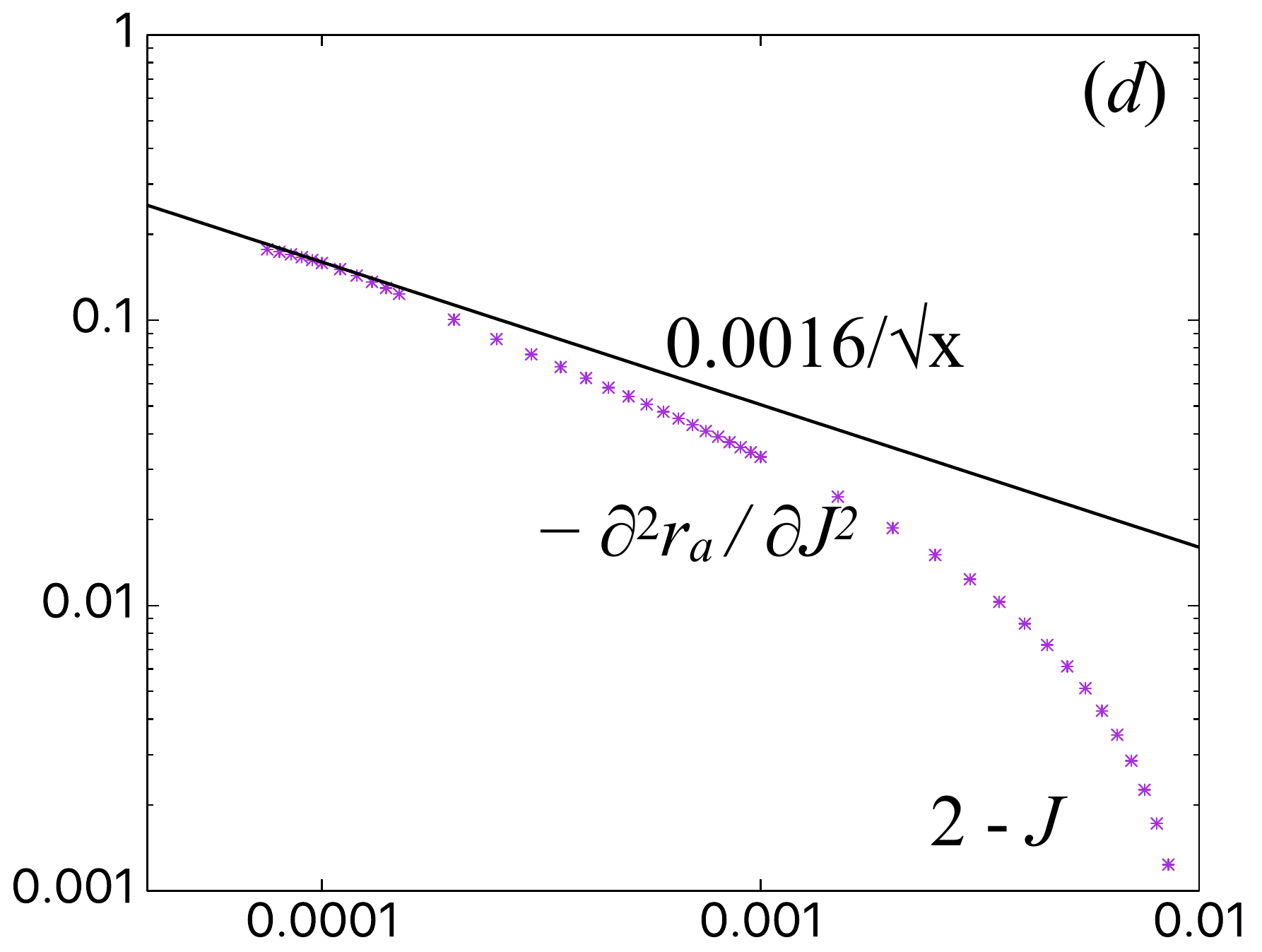}
\caption{Rate function $r_a$ and its two derivatives computed numerically from Eqs. (\ref{r_inf2}, \ref{d_r}) for quench of the parameter $J_3$  at $\beta=0.1$, $J_1=J_2=1$. For (a), (b), the pre-quench value $J_0$ is within the gapless phase, while for (c), (d) $J_0$ is within the gapped phase. In both cases the quantity $\partial^2 r_a/\partial J^2$ diverges at the phase boundary $J_0=2$. }
\end{figure}

\begin{figure}[h]
\includegraphics[scale=0.24]{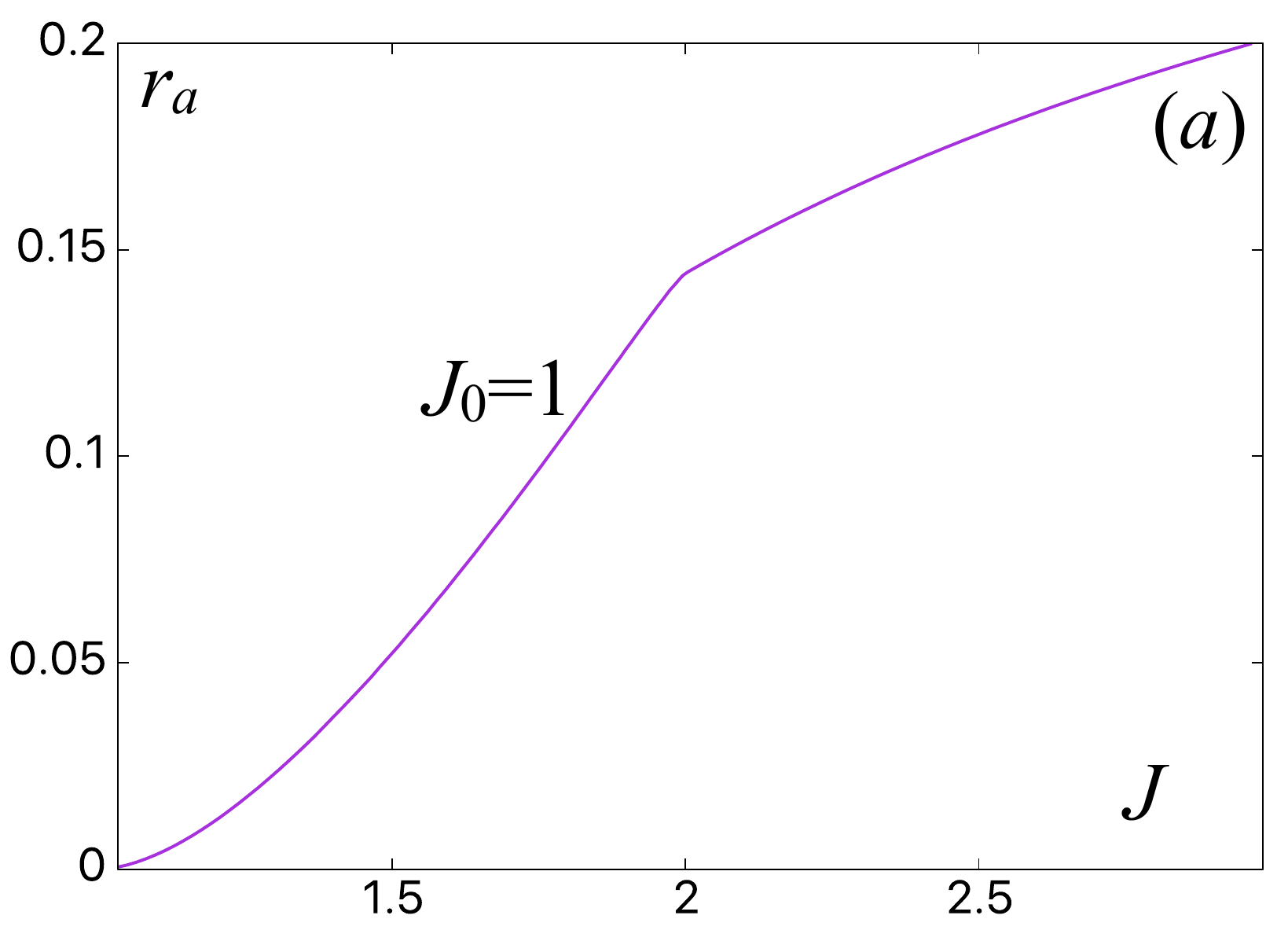}
\includegraphics[scale=0.24]{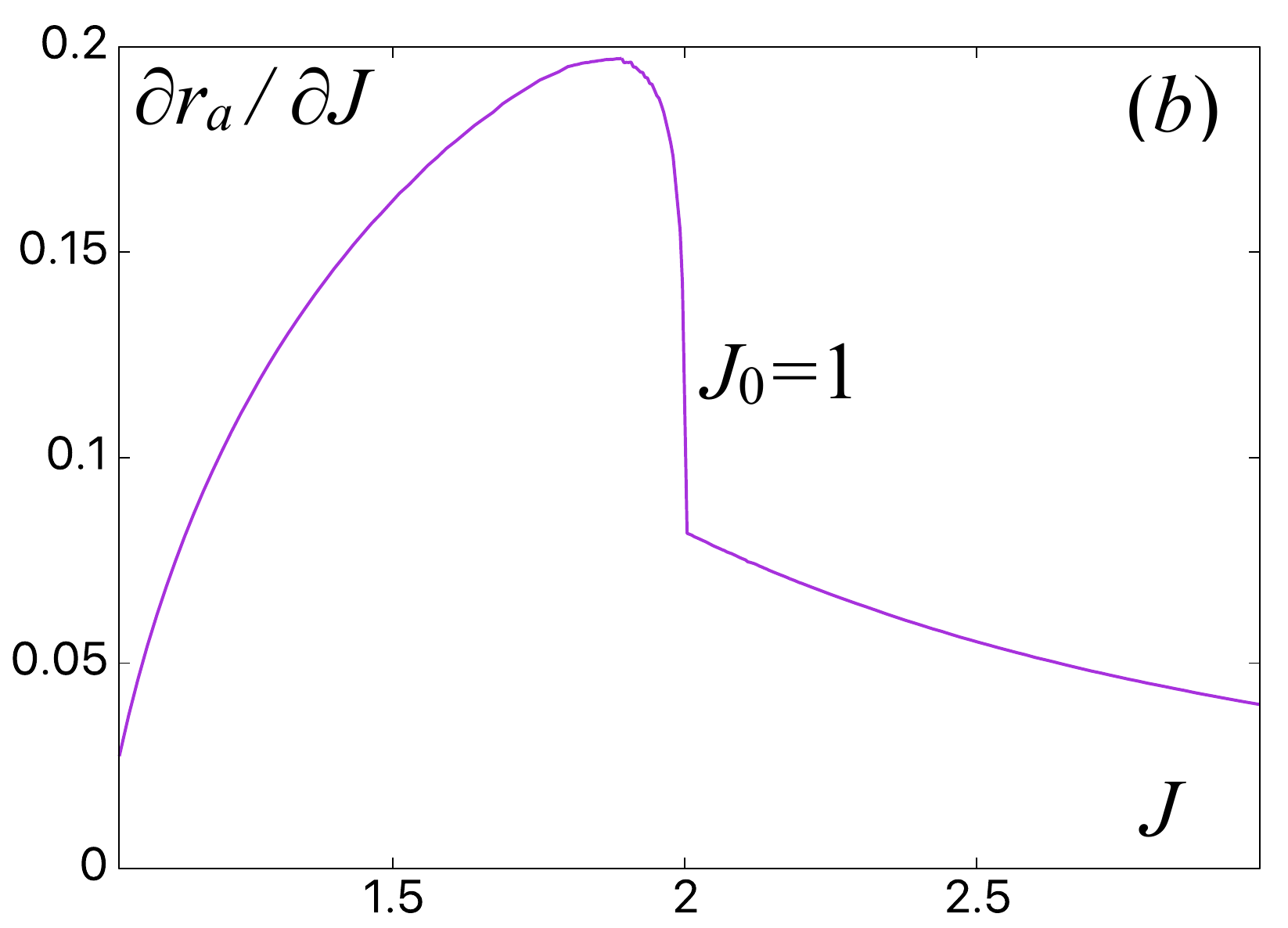}
\includegraphics[scale=0.24]{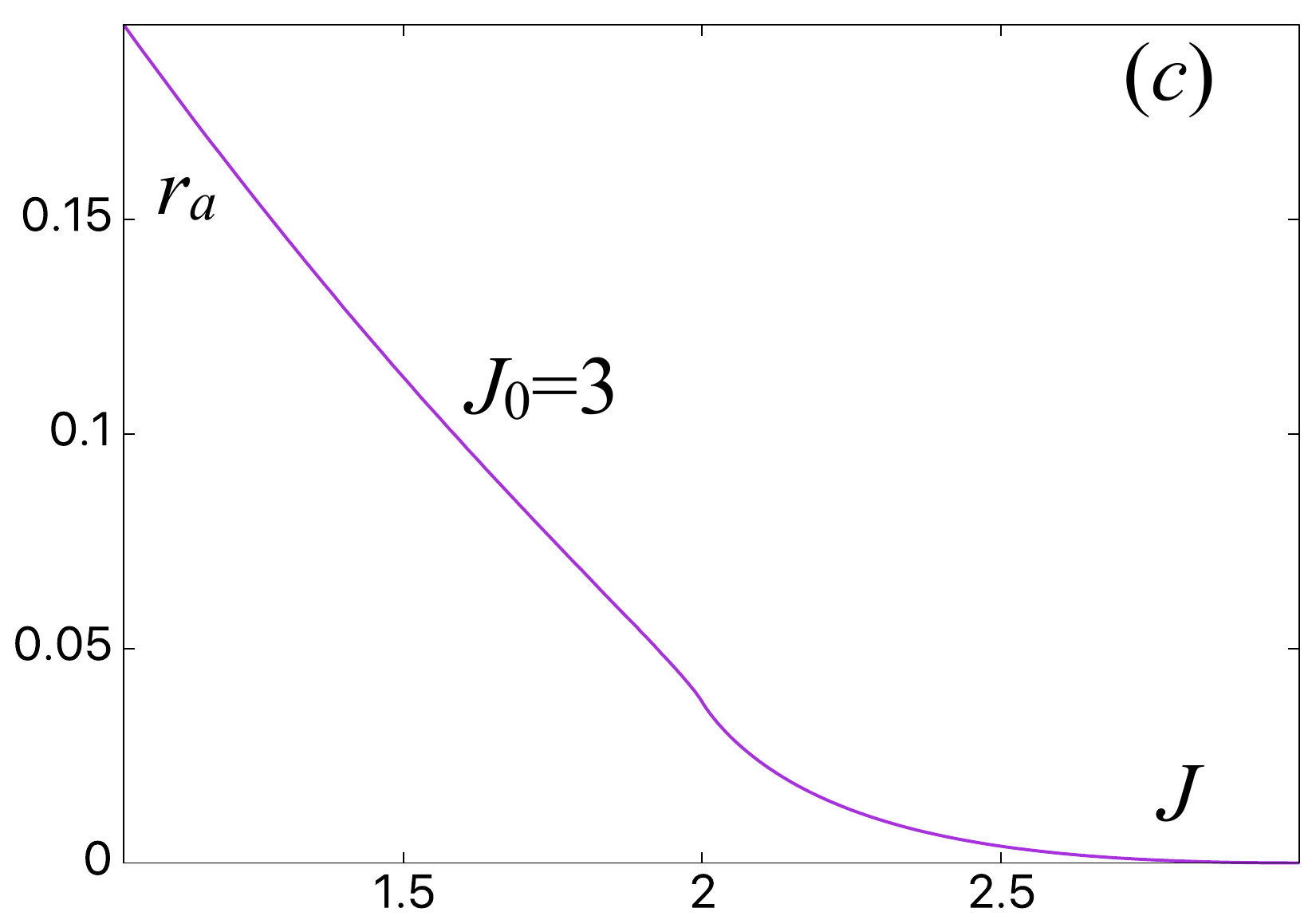}
\includegraphics[scale=0.24]{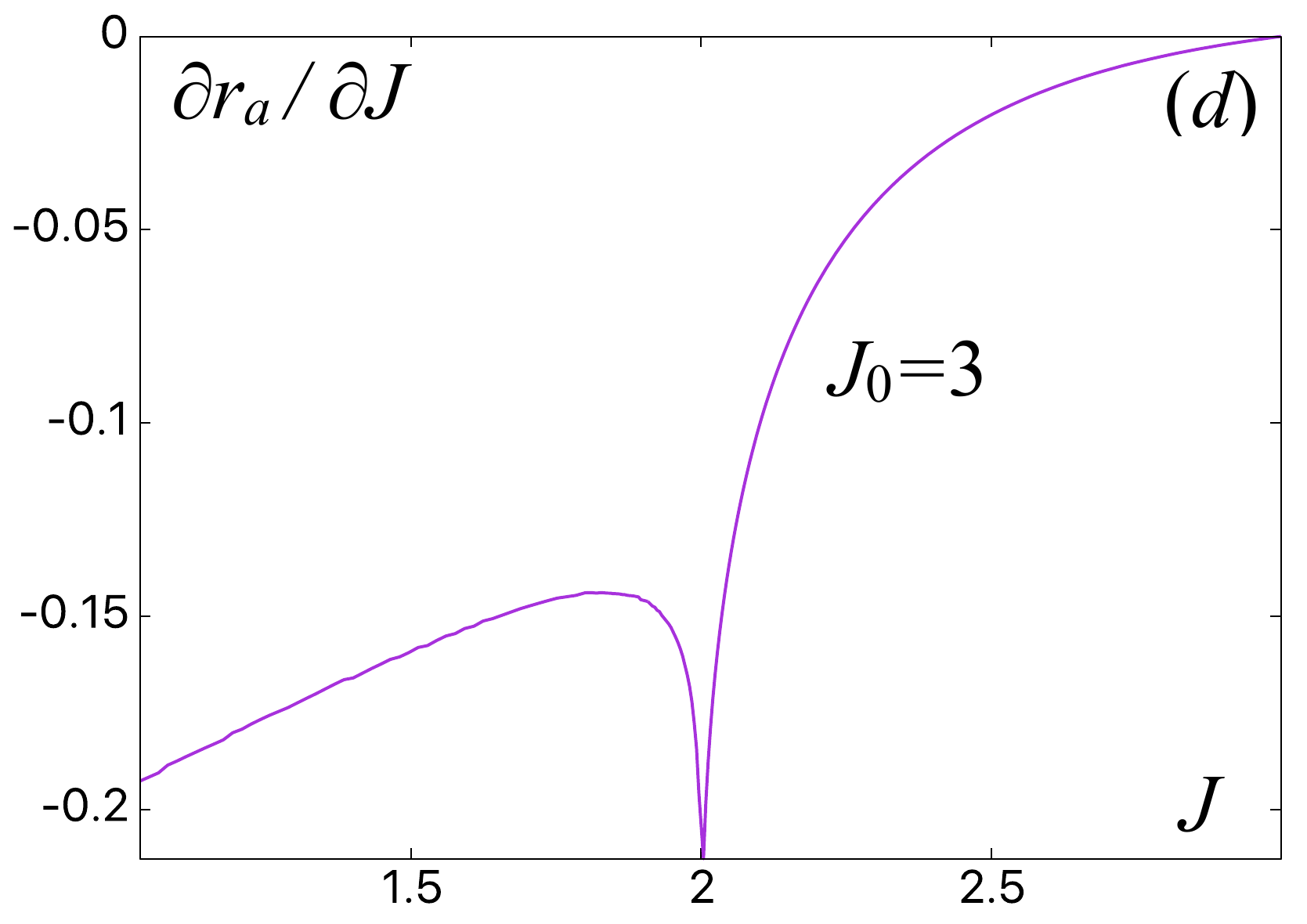}
\caption{Rate function $r_a$ and its two derivatives computed numerically from Eqs. (\ref{r_inf2}, \ref{d_r}) for quench of the parameter $J_3$  at $\beta=\infty$, $J_1=J_2=1$. For (a), (b), the pre-quench value $J_0$ is within the gapless phase, while for (c), (d)  $J_0$ is within the gapped phase. In both cases the quantity $\partial^2 r_a/\partial J^2$ diverges at the phase boundary $J_0=2$. }
\end{figure}

\newpage

\section{Results for two 3D Hamiltonians applicable to semimetals}

\begin{figure}[h]
\includegraphics[scale=0.3]{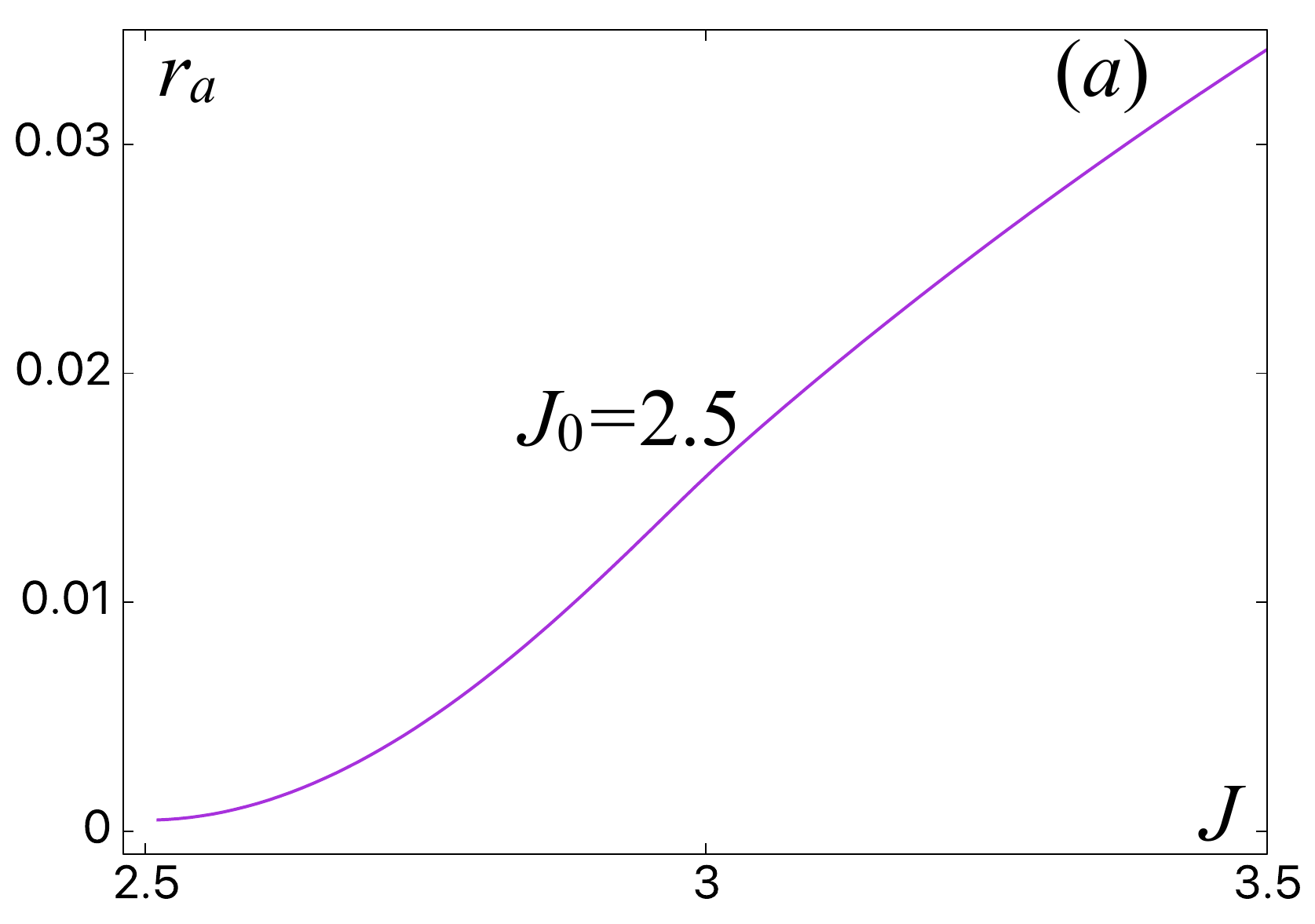}
\includegraphics[scale=0.3]{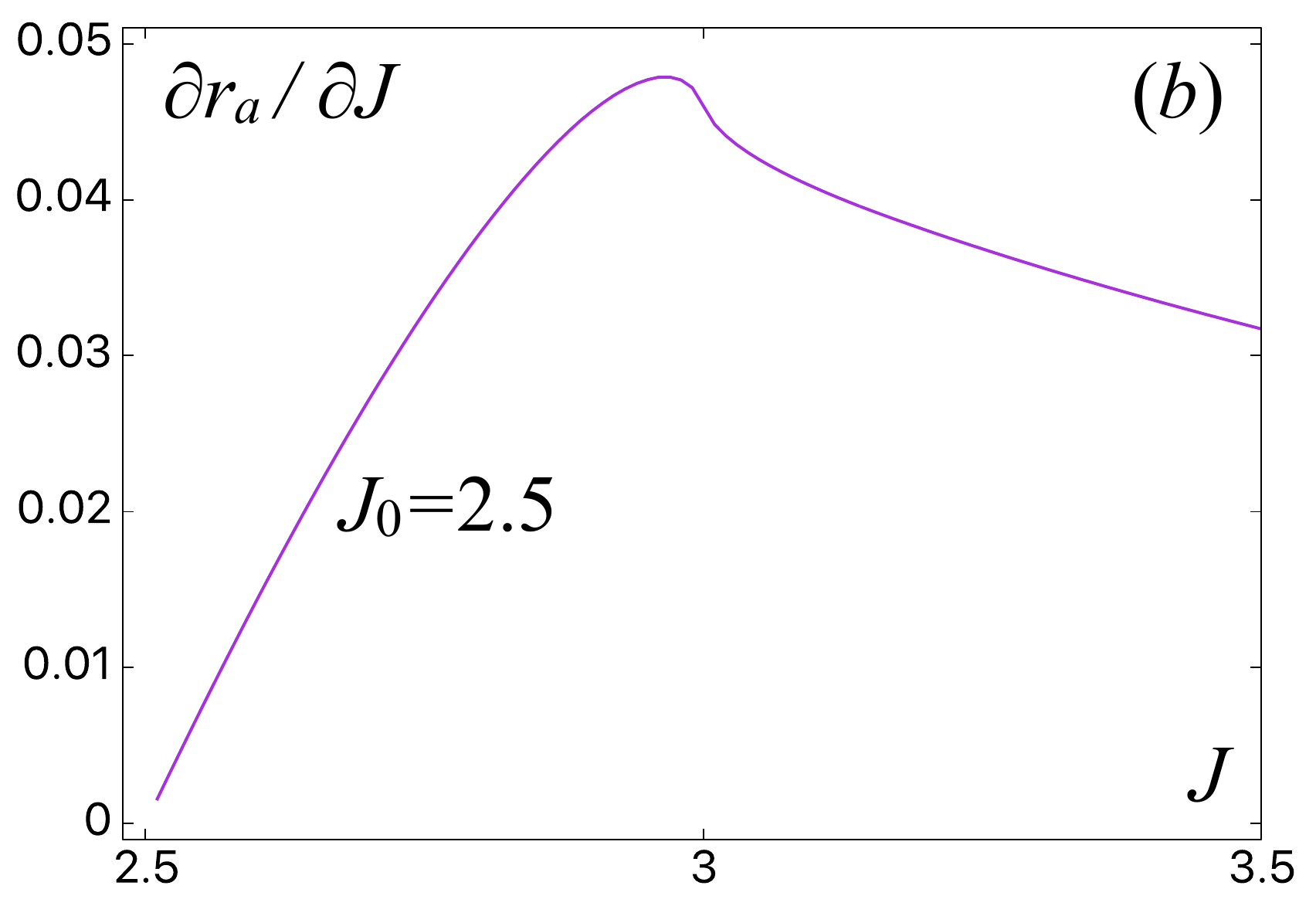}
\includegraphics[scale=0.3]{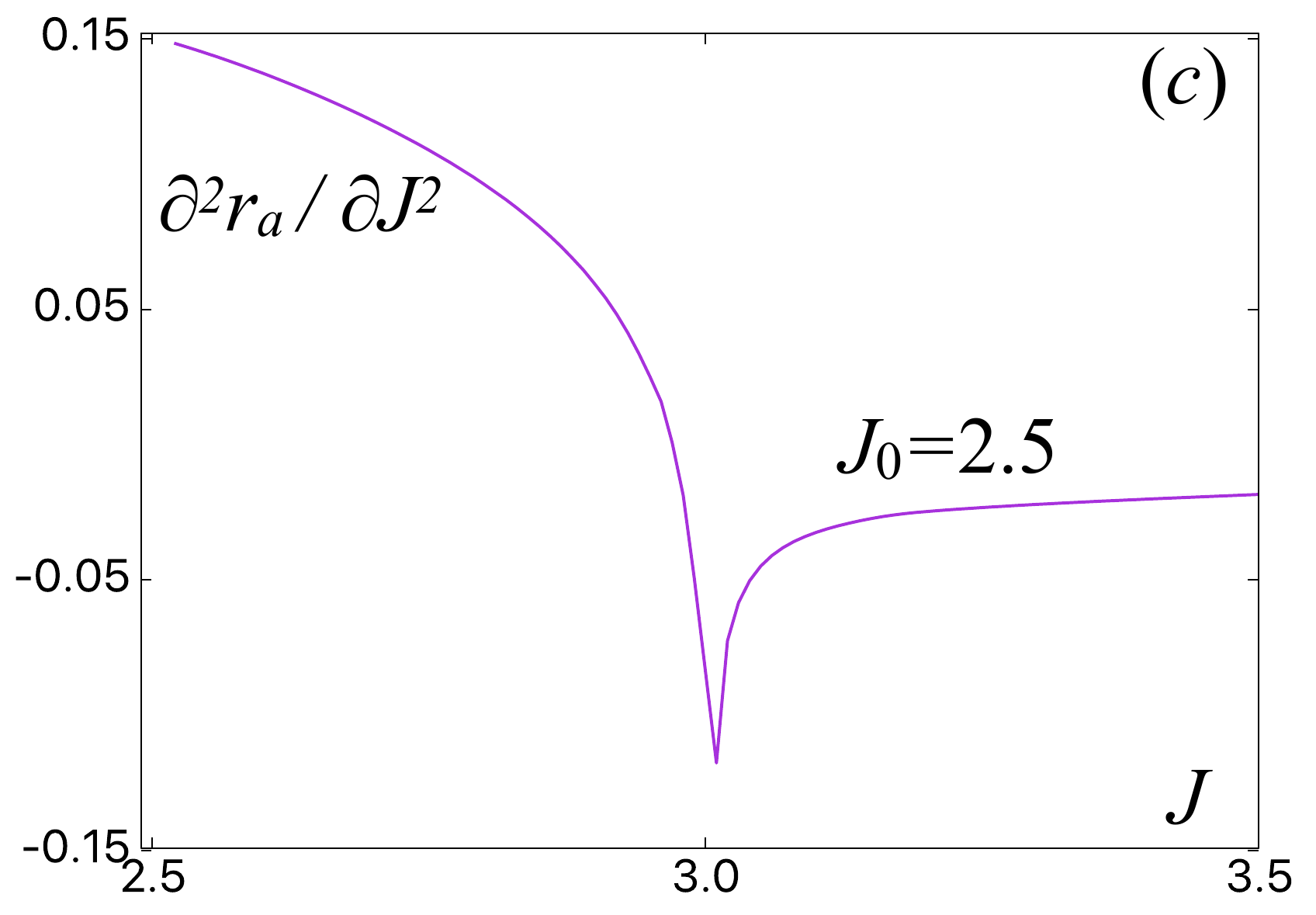}
\includegraphics[scale=0.3]{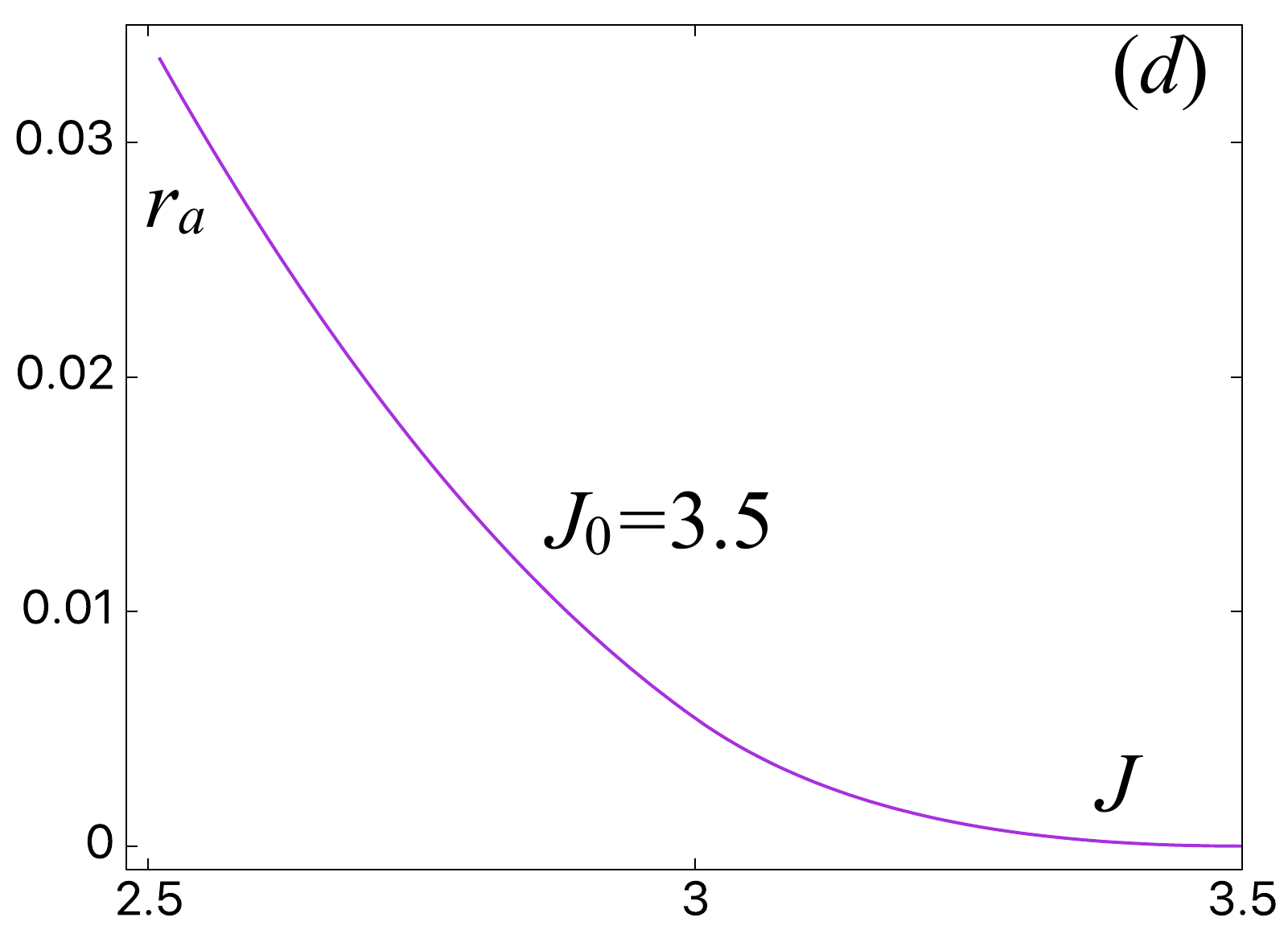}
\includegraphics[scale=0.3]{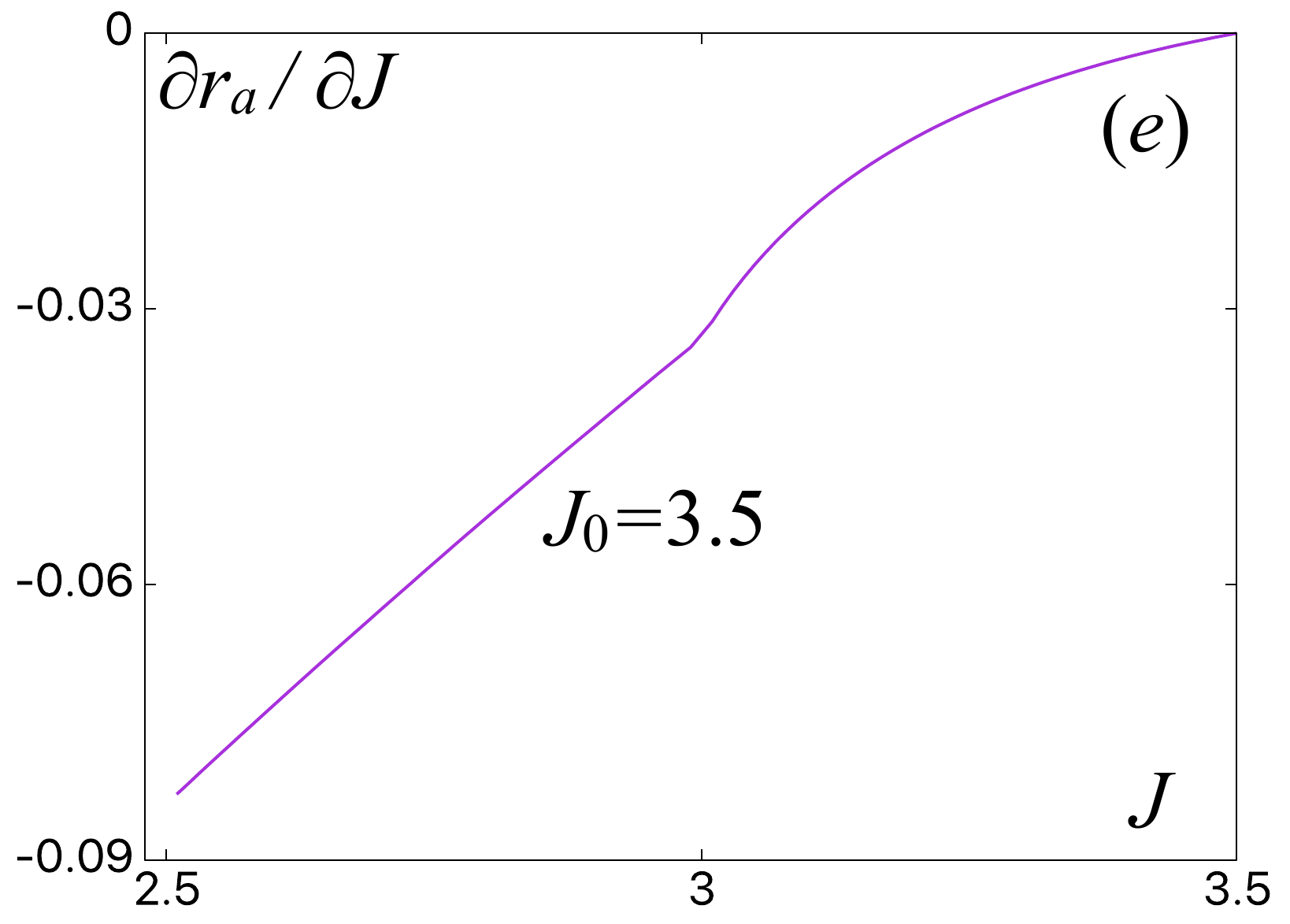}
\includegraphics[scale=0.3]{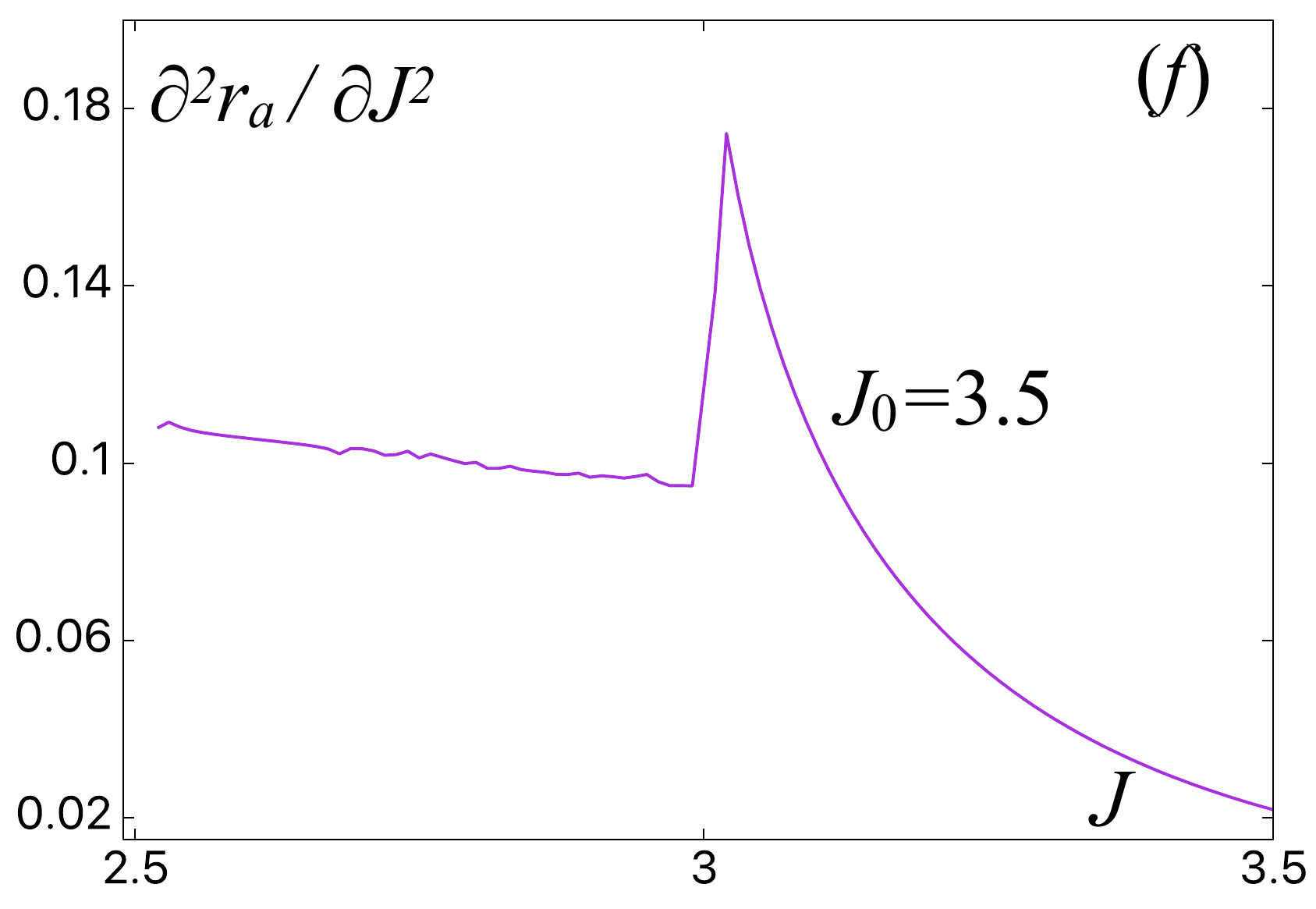}
\caption{Rate function $r_a$ and its two derivatives computed numerically from Eqs. (\ref{r_inf2}, \ref{d_r}) for quench of the parameter $J$  at $\beta=5$ in the case of Weyl semimetal Hamiltonian. For (a), (b) and (c), the pre-quench value is within the gapless phase, while for (d), (e) and (f) it is within the gapped phase. In both cases the quantity $\partial^2 r_a/\partial J^2$ diverges at the phase boundary $J_0=2$. }
\end{figure}

\begin{figure}[h]
\includegraphics[scale=0.3]{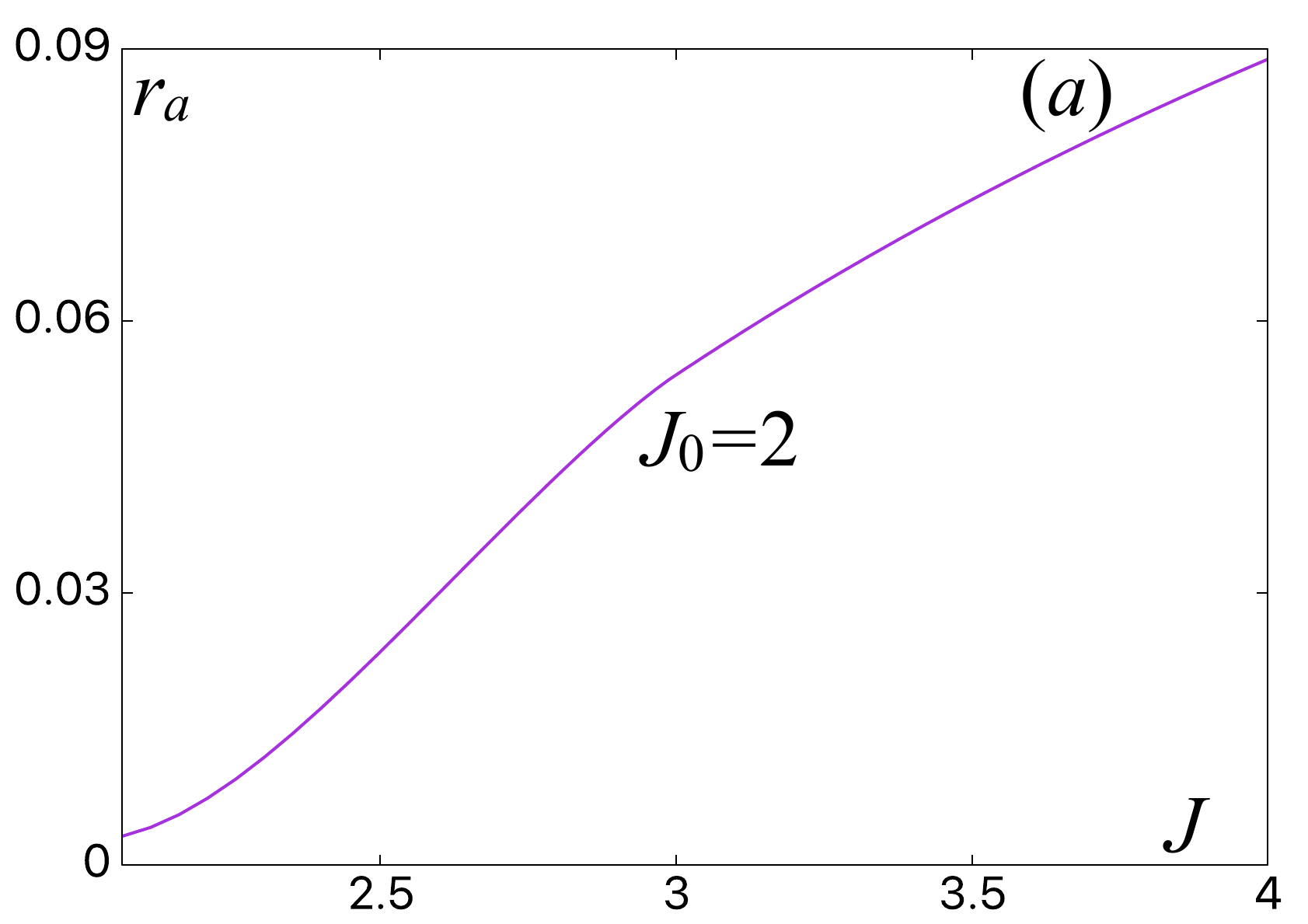}
\includegraphics[scale=0.3]{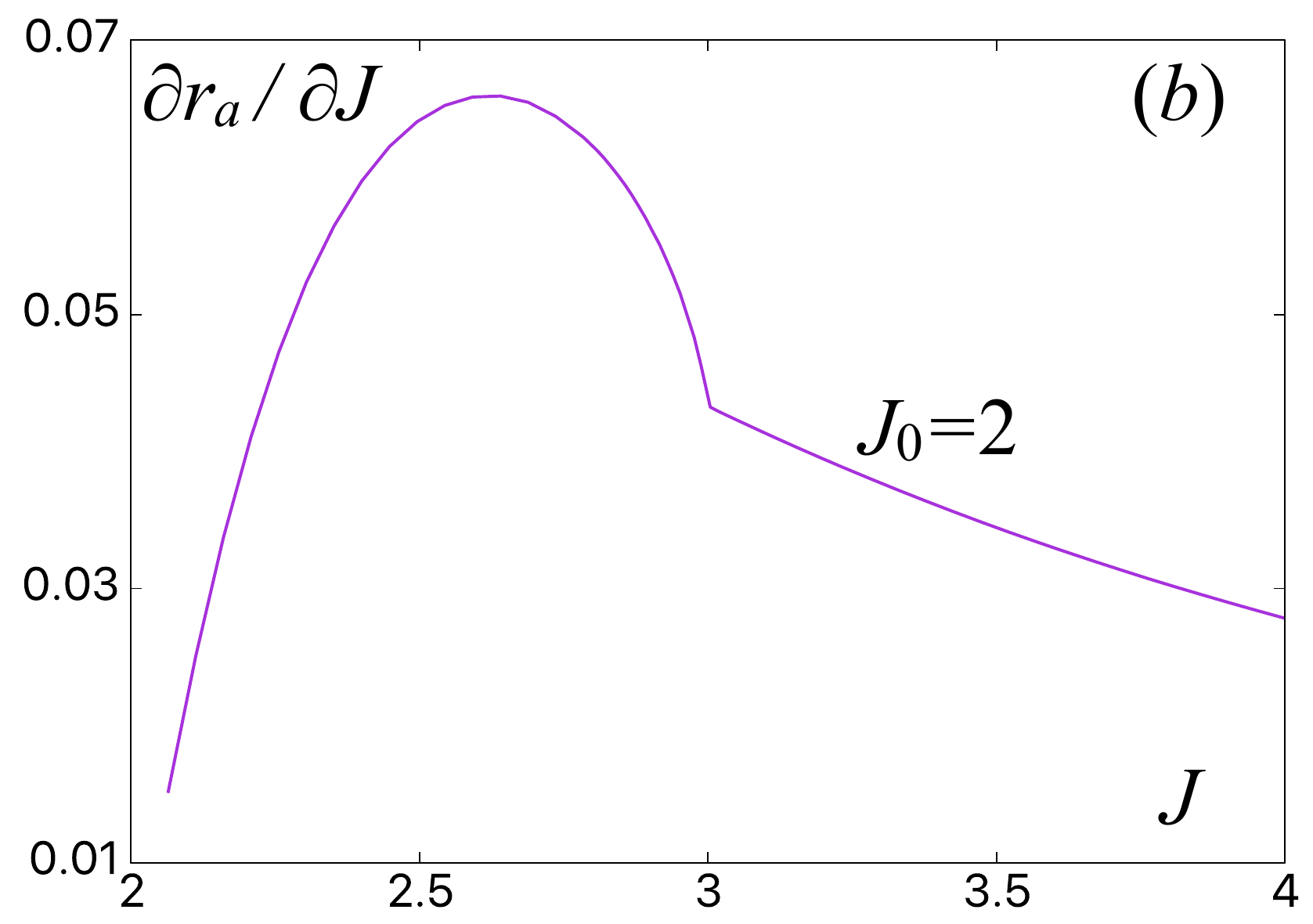}
\includegraphics[scale=0.3]{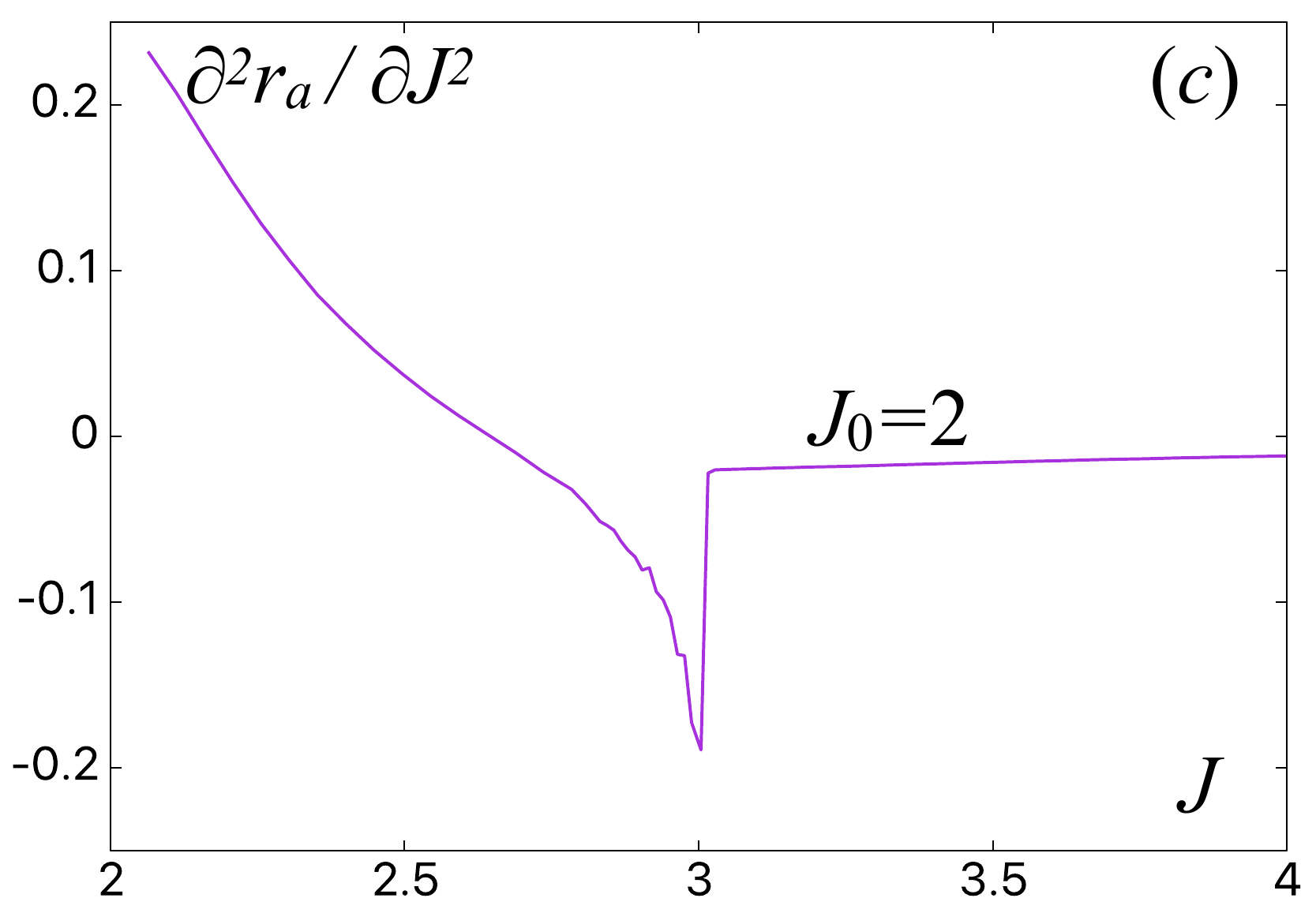}
\includegraphics[scale=0.3]{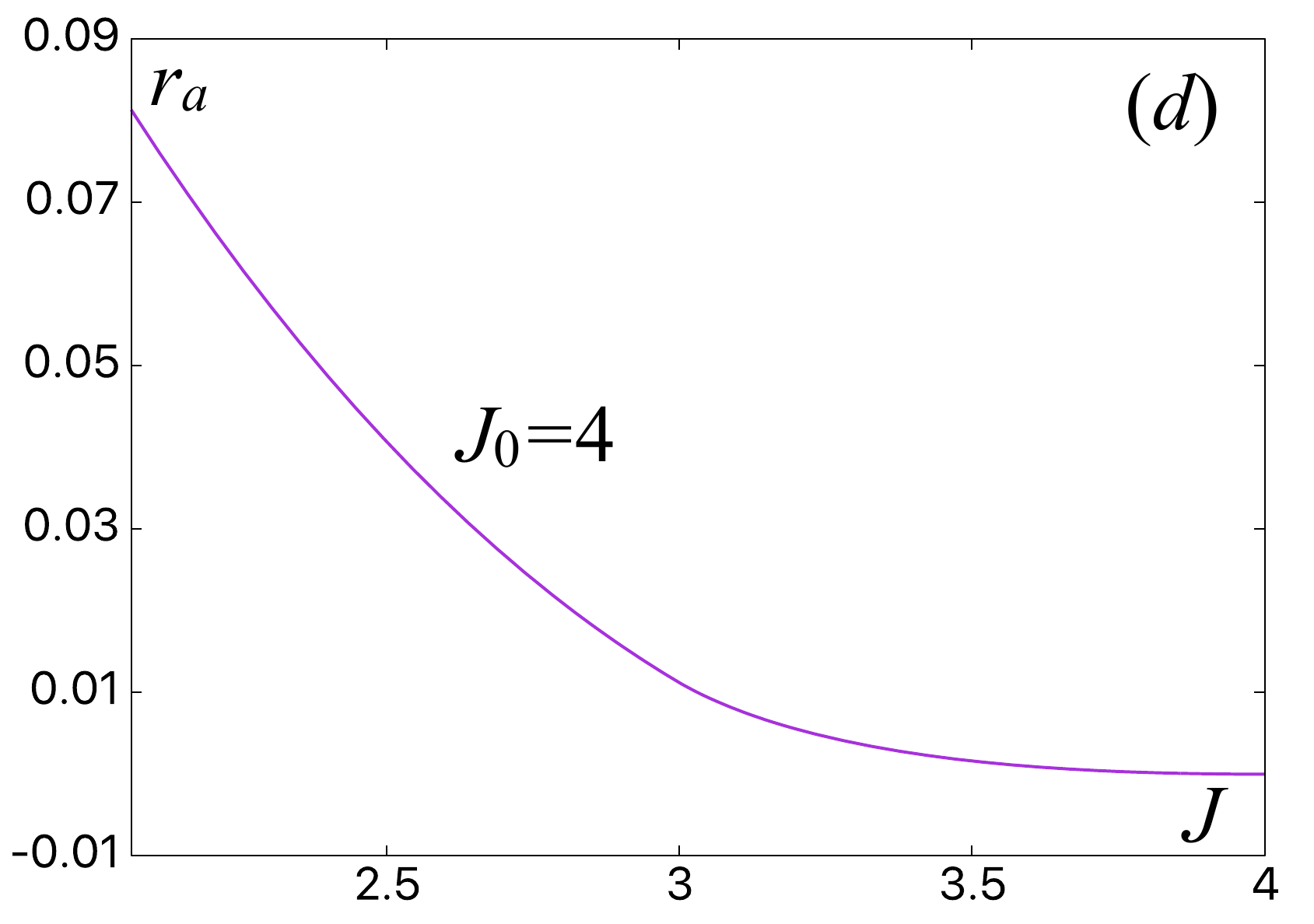}
\includegraphics[scale=0.3]{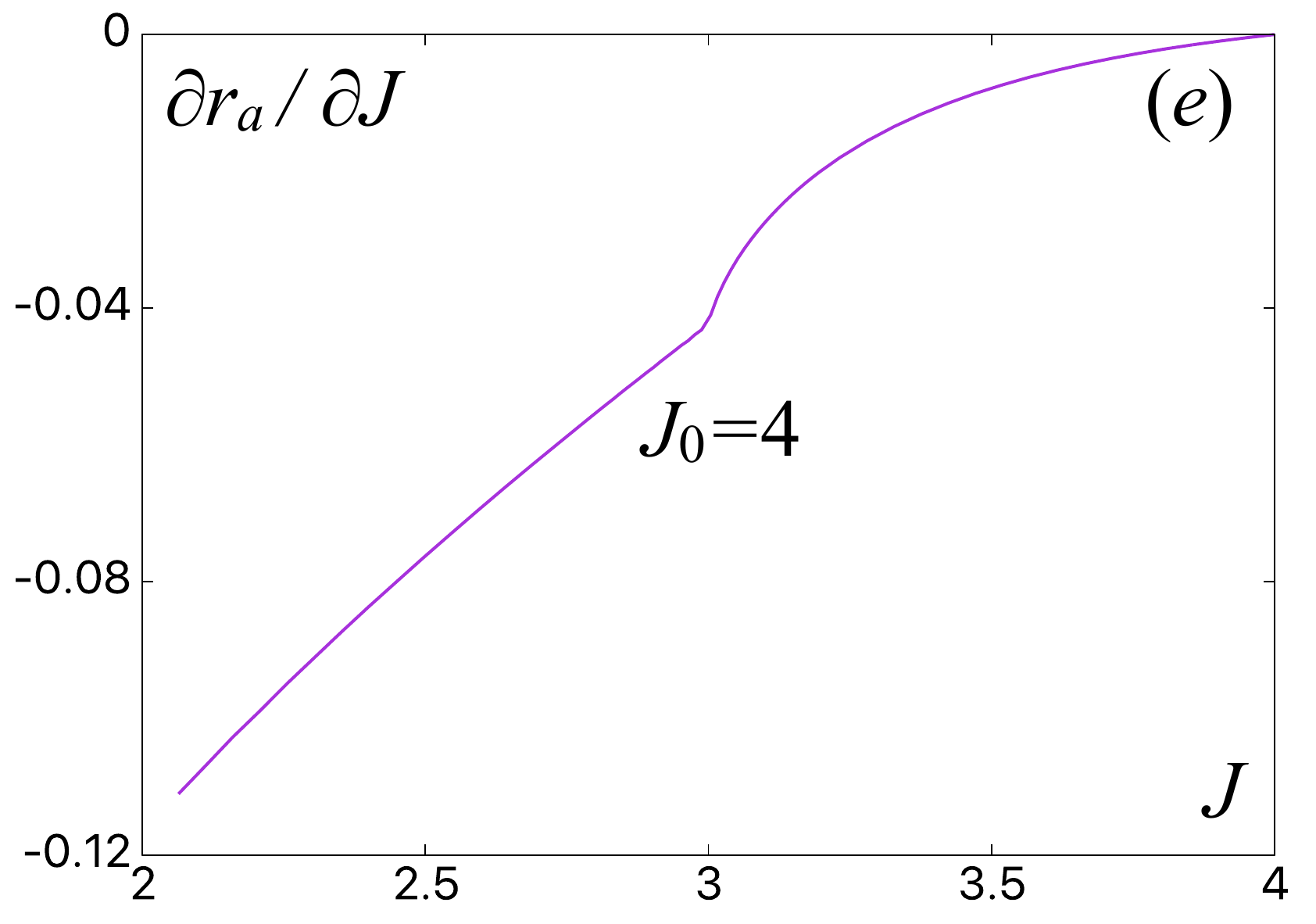}
\includegraphics[scale=0.3]{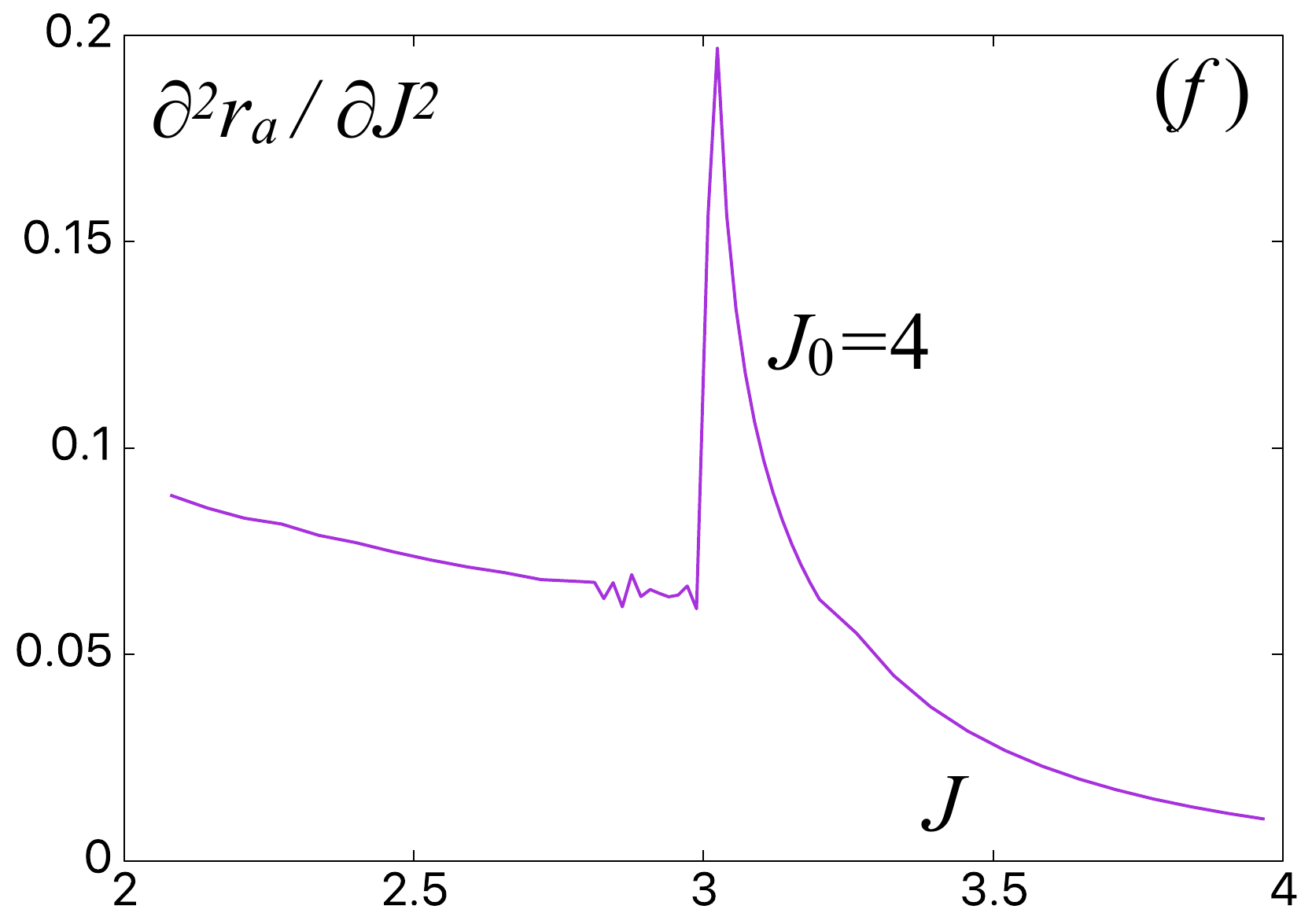}
\caption{Rate function $r_a$ and its two derivatives computed numerically from Eqs. (\ref{r_inf2}, \ref{d_r}) for quench of the parameter $J$  at $\beta=5$ in the case of 3D Hamiltonian with time reversal symmetry. For (a), (b) and (c), the pre-quench value is within the gapless phase, while for (d), (e) and (f) it is within the gapped phase. In both cases the quantity $\partial^2 r_a/\partial J^2$ diverges at the phase boundary $J_0=2$. }
\end{figure}


\bibliography{dmt}
\bibliographystyle{apsrev4-2}